# Molecular organization in the twist-bend nematic phase by resonant X-ray scattering at the Se K-edge and by SAXS, WAXS and GIXRD

W. D. Stevenson[1], Z. Ahmed[2], X. B. Zeng[1,*], C. Welch[2], G. Ungar[1,3] and G. H. Mehl[2,*]

[1] Department of Materials Science and Engineering, University of Sheffield, Sheffield S1 3JD, UK
[2] Department of Chemistry, University of Hull, Hull HU6 7RX, UK
[3] Department of Physics, Zhejiang Sci-Tech University, Hangzhou 310018, China

**Abstract:**

Using a novel Se-labelled dimer mixed with DTC5C7 aligned by magnetic field, the twist-bend nematic phase ($N_{tb}$) in dimers was studied by hard X-ray resonant scattering and by small and wide angle X-ray scattering (SAXS, WAXS). Resonant diffraction spots indicated a helix with a 9-12 nm pitch in the $N_{tb}$ phase. Unprecedentedly high helix orientation enabled deconvolution of global and local order parameters. This, combined with simultaneous resonant and non-resonant SAXS and WAXS data, allowed us to construct a modified model of the $N_{tb}$ phase matching twisted molecular conformations and the local heliconical director field.

The nematic-nematic liquid crystal (LC) transition, first reported in main chain LC polyethers [1],[2] and later in bent LC dimers [3]-[21], occurs where chemically linked mesogens are angled to each other (the bend angle) by a semi-flexible linker, such as an odd-numbered oligomethylene spacer. It has been proposed that the lower temperature nematic has a helical superstructure where the nematic director follows a heliconical path [22]. The helical structure is considered to be stabilised by local twist and bend elastic distortions; hence the common name 'twist bend nematic', or '$N_{tb}$' phase [23],[24]. In current theory the mesogens glide fluidly along the helical axis, possess no long range positional order and complete a full rotation in just a few molecular lengths.

Topological characterisation methods, e.g. opto-electric studies [3],[4], have provided support for the helical theory of the $N_{tb}$ phase. A nano-scale pitch has been proposed based on AFM [5] and freeze fracture TEM [6],[7]; though in both cases the imaging was performed below the stability range of the $N_{tb}$ phase, and doubts remained as to whether the surface imaging was representative of the bulk. Recently helical periodicity within the $N_{tb}$ phase has been directly identified in a bulk sample by resonant X-ray scattering at the carbon absorption edge (K-edge) [8]. A resonant Bragg ring was observed in the $N_{tb}$ phase originating from a 8–10 nm long helical pitch. Such observation of orientational periodicity is possible because at the absorption edge of an atom, X-rays scattering becomes sensitive to the symmetry of its bound orbitals, and the atomic form factor changes from scalar to a complex tensor [25]-[27].

Although the helical pitch has now been confirmed by carbon RXS, many questions about the $N_{tb}$ phase remain unanswered. The three principal subjects of debate regard the orientational order parameter, the manner in which the molecules pack into the helix and whether or not the molecules are locally layered. However, carbon RXS is not able to provide answers to such questions, because the low energy 'soft' X-rays (~0.3 keV) required for carbon resonance imposes severe experimental constraints, such as a fully windowless beamline in high vacuum and the sample held between thin fragile substrates, not allowing alignment by external fields [8]. Moreover, soft X-ray scattering allows access to only a limited range of reciprocal space, excluding scattering at larger wavevectors $q$ that holds information on molecular conformation and packing.

Here we avoid the drawbacks of carbon RXS by studying the $N_{tb}$ phase using harder X-rays at the selenium K-edge (12.658 keV). For this purpose a selenoether dimer was specially synthesised. In addition to resonant diffraction from the heliconocal $N_{tb}$ structure, we simultaneously record non-resonant scattering at larger $q$. We align the $N_{tb}$ phase using magnetic field and, for the first time, obtain direct information on the orientation of the helical axis and, simultaneously, that of the molecules. This enabled us to obtain unprecedented details about molecular conformation and packing in the $N_{tb}$ phase. Combined resonant and non-resonant SAXS and WAXS data on the perfectly aligned sample, together with grazing incidence X-ray diffraction (GIXRD), allowed us to construct a modified model of the $N_{tb}$ phase. This includes twisted molecular conformations matching the shape of the local heliconical director field, as well as local molecular layering. We also present the first quantitative data on correlation lengths/helical domain size in bulk $N_{tb}$.

The Se-labelled compound DTSe (Figure 1a) was synthesised as described in supporting information (SI). DTSe does not exhibit the $N_{tb}$ phase, but does so when mixed with the previously studied compound DTC5C7 (Figure 1a) [9],[10]. The miscibility of the two compounds was examined qualitatively by bringing them into contact under crossed polarisers. A composition gradient was formed, which was most evident at 114°C. In Figure

1b the N phase in the centre separates the $N_{tb}$ phase of the DTC5C7-rich region on the left and the Smectic-C (SmC) phase of DTSe-rich region on the right.

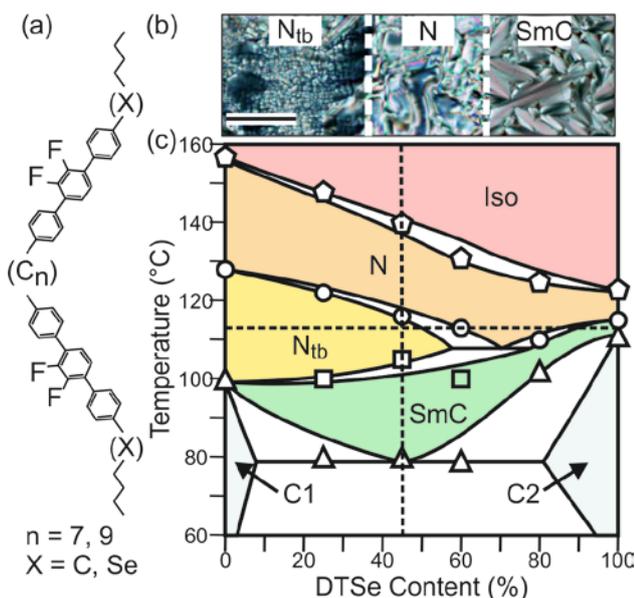

Figure 1. (a) The molecular structure of DTC5C7 (n= 7, X= CH$_2$) and DTSe (n= 9, X= Se). (b) Optical textures observed between cross polarizers at 114°C for a contact preparation, showing the change from $N_{tb}$, to N, to SmC phase with increasing concentration of DTSe in the binary mixture. (c) Phase diagram of DTC5C7 and DTSe. Experimental points are DSC peak temperatures on first heating. C1= DTC5C7 crystal, C2= DTSe crystal. The dashed vertical line indicates Se45, and the horizontal line 114°C.

The binary phase diagram (Figure 1c) was determined by differential scanning calorimetry (DSC) and polarised optical microscopy (POM). The diagram shows a stable $N_{tb}$ region between 0 and ca 60 mol% DTSe. For our RXS investigation we selected a mixture containing 45 mol% DTSe, henceforth referred to as Se45.

The RXS investigation of Se45 was performed on beam-line I22 of the Diamond Light Source. The sample was held in a 1 mm glass capillary, positioned perpendicular to a 1T magnetic field. The sample was heated and cooled using nitrogen gas. Se45 was investigated in each of its phases using X-ray energies above and below the Se K-edge ($E_r$ = 12.658 keV). Two Pilatus 2M detectors were simultaneously employed to cover the range of ~ 0.3 nm$^{-1}$ ≤ $q$ ≤ 15 nm$^{-1}$. Non-resonant small and wide-angle X-ray scattering (SAXS and WAXS) experiments were carried out at station BM28 (ESRF) using a MAR165 CCD camera.

The sample was first cooled from the isotropic to the $N_{tb}$ phase (105°C) where incident energy ($E_i$) was increased in increments of 5 eV from 12.608 to 12.708 keV. Throughout this energy range two broad SAXS maxima were observed at $q$ = 1.44 and 3.16 nm$^{-1}$, corresponding to average spacings of 4.4 and 2.0 nm (Figure S4). These spacings are comparable to the average dimer length (4.5nm) and centre-to-centre mesogen distance (2.1nm) within Se45. A non-resonant WAXS maximum at $q$ = 13.7 nm$^{-1}$ corresponds to a d-spacing of 0.46 nm, suggesting an average lateral intermolecular distance of 0.51 nm [28]. However at X-ray energies satisfying $|E_i - E_r|$ ≤ 5eV, an additional Bragg peak was observed at $q$ = 0.62 nm$^{-1}$, attributed to resonant diffraction from a helical pitch $p$ of 10.1 nm (Figure S4). The resonant peak was strongest at $E_i = E_r$.

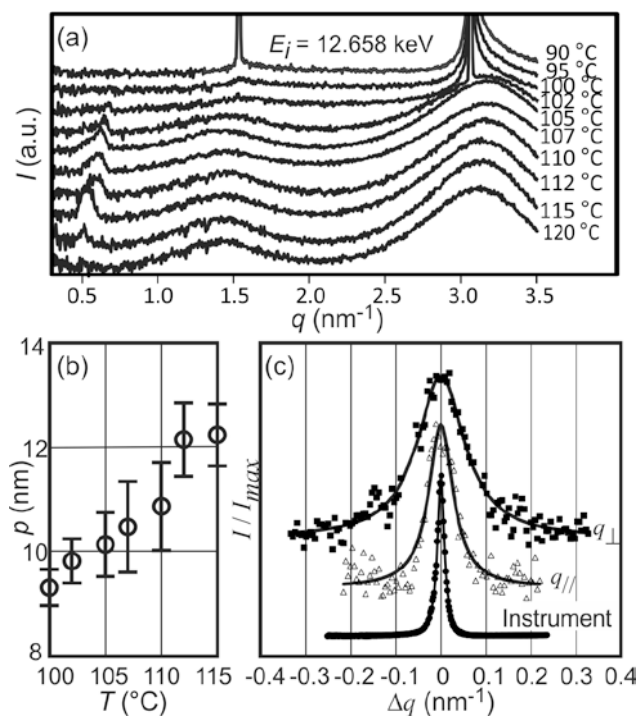

Figure 2. (a) The resonant diffraction peak is only observed in the $N_{tb}$ phase of Se45. (b) The helical pitch $p$ of $N_{tb}$ phase decreases with cooling. (c) Fit of a Lorentzian profile to the resonant diffraction peak of the $N_{tb}$ phase, recorded at 112°C.

Temperature dependence of the resonance effect was investigated by cooling the sample from isotropic through each of its LC phases with $E_i = E_r$. The resonant peak was observed exclusively in the $N_{tb}$ phase and at the biphasic boundaries. As shown in Figure 2a, the resonant peak appeared at the onset of the $N_{tb}$ phase (115°C), reaching maximum intensity at 112°C, with $p$ = 12.1 nm. With decreasing temperature the resonant peak shifted to larger $q$, indicating a reducing pitch length (Figure 2b). The radial full width at half maximum (FWHM) of the peak initially broadened, indicating an increased distribution of $p$, before sharpening again at the onset of the SmC phase. The resonance effect was lost with further cooling into the SmC phase. These findings are summarised numerically in Table S2 (SI).

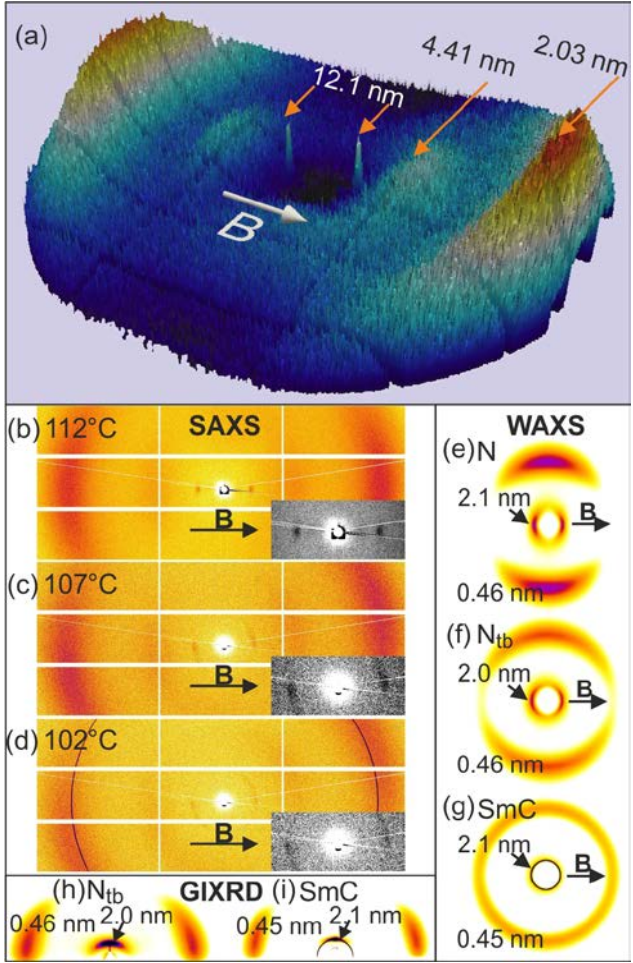

Figure 3. (a) Surface plot of the SAXS diffractogram of the $N_{tb}$ phase of Se45 at 112°C aligned in magnetic field, $E_i$ = 12.658 keV. (b)(c)(d) 2D SAXS patterns of the $N_{tb}$ phase at 112°C, 107°C and 102°C, with zoom-in of the sharp resonant SAXS peaks. (e-g) Transmission WAXS patterns of Se45 in the (e) N, (f) $N_{tb}$ and (g) SmC phase. (h-i) Wide-angle GIXRD patterns of Se45 in the (h) $N_{tb}$ and (i) SmC phase; smectic layers are parallel to the horizontal substrate.

At 112°C the sharp resonant peaks were exactly on the meridian (Figure 3a,b), defined as the axis through the beam centre, parallel to the magnetic field. This indicates that the helical axis of the bulk sample was highly oriented with the field. The diffuse SAXS peaks were azimuthally spread but were also centred on the meridian. With the $N_{tb}$ phase highly aligned, the azimuthal breadth of the SAXS maxima reveals that the short-range layers are not all perpendicular to the helical axis.

On cooling from 112°C the intensity maxima of all three SAXS peaks simultaneously shifted away from the meridian; the lower the temperature, the larger the shift – see Figures 3b-d. The resonant peak began to broaden azimuthally, indicating decreasing orientational order. Phase orientation was lost entirely in the SmC phase – see the sharp Bragg ring in Figure 3d. The lamellar phase is assumed to be SmC because the layer height (4.12 nm) is shorter than the length of both dimers (tilt angle ~24°). The WAXS maxima similarly rotated and spread azimuthally, before becoming a diffuse halo in the SmC (Figure 2e-g).

The temperature dependency of the pitch length was reported by C. Zhu et al [8] who measured $p$ of CB7CB to vary between 8 and 10nm. In their work the reflection was usually split covering a wider $q$-range, possibly due to the surface confinement effect. The increased $p$ of Se45 (9 – 12nm) is attributed to the larger molecular lengths of DTC5C7 and DTSe. At 112°C the azimuthal FWHM of the resonant peak was ~12°, indicating that the orientational order parameter $S$ of the global helical axis must be close to 1 (($3\cos6°-1)/2=0.992$). Assuming a perfect orientational order of the helical axis, the intensity distribution of the resonant peak can be attributed to structural correlations parallel and perpendicular to the helical axis. In a nematic LC one may assume that correlation decays exponentially with distance $r$, i.e. $\propto \exp(-r/\xi)$. Under this assumption the scattered intensity in reciprocal space is well reproduced by Lorentzian profiles of the form $I_{max}/(1+\Delta q_{//}^2\xi_{//}^2+\Delta q_\perp^2\xi_\perp^2)$ [29]-[31], where $\xi_{//}$ and $\xi_\perp$ are the correlation lengths parallel and perpendicular to the helical axis. By least squares fit of a Lorentzian profile to the intensity around the resonant peak at 112°C ($I$ vs $\Delta q_{//}$ and $I$ vs $\Delta q_\perp$), and after corrections for instrumental broadening (Figure 2c), we obtain $\xi_{//}$ = 33 nm and $\xi_\perp$ = 18 nm. Alternatively, using the FWHM of the Lorentzian fit ($\Delta q=2/\xi$) to the resonant peak, the Scherrer equation $\tau = 2K\pi/\Delta q$ can be used to estimate the size of the coherent heliconical domain. If the shape factor ($K$) is set to unity, then $\tau = \pi\xi$ and the longitudinal and transverse helical domain sizes are estimated to be 104 nm (24 molecular lengths) and 57 nm (112 molecular widths). Note that both values are underestimates: along the meridian we have ignored the small distribution of helical pitch lengths, while in the transverse direction we have assumed perfect orientational order of the helical axis. At temperatures below 112°C, $\xi_\perp$ can no longer be determined due to the contribution of $S<<1$ to azimuthal peak broadening. Similarly the increasing distribution of $p$ with reducing temperature prevents accurate measurement of $\xi_{//}$. It appears that $\xi_{//}$ reduces as the $N_{tb}$ phase progresses, until reaching the onset of the SmC phase, where the quasi-long-range positional order sets in [32].

As discussed above, both broad SAXS peaks arise from short range correlations of dimers and mesogens. At temperatures close to the N-$N_{tb}$ boundary, the inter-mesogen correlation length along the helical axis can be estimated from $\Delta q_{//d2}$ (Table S2) as ~1.7 nm. Furthermore the two broad SAXS peaks remained centred on the helical axis throughout the temperature range, despite the reducing effect of the magnetic field. The diffuse SAXS peaks are evidence of local longitudinal register between dimers, which can be regarded as localised layering, usually driven by π-π-interactions between aromatic mesogens and often referred to as cybotactic nematic correlations [28].

If the mesogens are tilted by an angle $\theta$ with respect to the helical axis, a second order parameter $S_1=(3\cos\theta-1)/2$ is required to describe the orientational distribution within each helical domain. The overall orientational order parameter $S$ of the $N_{tb}$ phase, measureable by classical means, is therefore $S_0 \cdot S_1$, which quantifies the orientational distribution of the helical domains ($S_0$) and the average tilt angle of the mesogens within each domain ($S_1$). We have shown that $S_0$ decreases with temperature in the $N_{tb}$ phase, even in the case of perfect helical alignment ($S_0=1$), $S_1$ will be small as $\theta >> 0°$ (see below). This imposes limitations on $S=S_0 \cdot S_1$, which may account for the low orientational order parameter (<P2> ~ 0.4 – 0.6) often reported within the $N_{tb}$ phase of bent dimers [11]-[15],[16].

Based on the above findings we propose that the molecules adopt a conformation resembling a helical segment (Figure 4a-e), thereby enabling the mesogens to follow the heliconical path of the local $N_{tb}$ director field. The overall structure is depicted in Figure 4g, which also illustrates the local layering responsible for the diffuse SAXS peaks.

The local director field along the helical axis can be defined mathematically as:

$$\mathbf{n_0} = \sin\theta\cos(2\pi z/p)\,\mathbf{i} + \sin\theta\sin(2\pi z/p)\,\mathbf{j} + \cos\theta\,\mathbf{k} \qquad (1)$$

Here $z$ is the height along the helical axis, $\theta$ is the tilt angle of the mesogens and $p$ the pitch length of the resulting helix. On progression along $\mathbf{z}$, the dimers can be imagined to wrap around a cylinder, such that each successive mesogen is rotated about the helical axis by $2\pi h/p$ (Figure 4d-f). $h$ denotes the cybotactic layer height, where $h \cong d2$ (Table S2). The tilt angle of the mesogens can also be calculated using the relation:

$$h = l\cos(\theta) \qquad (2)$$

where $l$ is the contour length between the centres of the two mesogens of each dimer (~2.3nm) (SI). As shown in Table S2, the value of $h$ is almost invariant with temperature (~2nm), suggesting a constant tilt angle of 29±1° in the $N_{tb}$ phase. This tilt angle is larger than that obtained from birefringence measurements on DTC5C7 [9], but broadly in line with those discussed for CBCnCB systems [12]; it is also slightly larger than the dimer tilt angle estimated for the SmC phase. A smaller tilt in the SmC is in fact suggested by GIXRD experiments on oriented thin films of Se45 (Figures 3h,i).

Using the experimental values of the rotation and tilt angles, one may also calculate the bend angle between the two mesogens of a dimer. If the two rod-like mesogens are respectively defined by unit vectors $\mathbf{n_1}$ and $\mathbf{n_2}$, it can be shown geometrically (SI) that:

$$\sin(\phi/2) = \sin(\theta) \cdot \sin(\pi h/p) \qquad (3)$$

where $\phi$ is the exterior (bend) angle between them (for a linear dimer $\phi=0$). At 112°C $\phi$ is calculated as 29°, which increases to 38° on cooling to 102°C. In both instances $\phi$ is surprisingly low and much smaller than the bend angle imposed by an all-trans spacer (>60°).

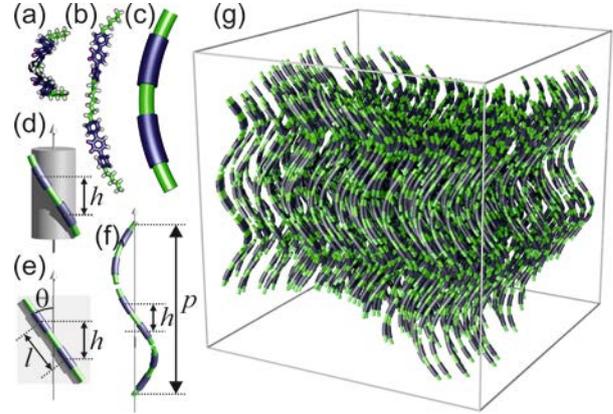

Figure 4. Model of the $N_{tb}$ phase. (a,b) A bent dimer adopts the shape of a segment of a helix; (a) projection along spacer axis, (b) side view. (c) Representation of the molecule as a helical segment. (d) A dimer wraps around a cylinder. $h$ = local layer height. (e) An "unwrapped" dimer. $\theta$ = tilt angle, $l$ = contour length between centres of mesogens. (f) Local arrangement of dimers along the helical axis. $p$ = pitch. (g) $N_{tb}$ phase assembled from molecules with conformations approximating helical segments; local layering is also shown.

The results suggest that in order to follow a heliconical path the molecules must adopt a higher energy conformation in which the arms are twisted about the spacer axis, and closer to parallel than in the minimum energy conformation. Molecular modelling (SI, Section 6) has shown that this can be achieved by changing the torsion angles of C-C bonds in the spacer, such as the molecule shown in Figures 4a,b. In the $N_{tb}$ phase it is likely that this distortion energy is compensated by the entropy associated with the retained positional freedom of the dimers. The increasing $\phi$ on cooling is therefore expected, as lower energy conformations are preferred at lower temperatures. Furthermore equation (3) shows that $p \propto 1/\phi$. On cooling, as $\phi$ increases the helical pitch $p$ is reduced. On further cooling, when the translational entropy gain becomes insufficient to outweigh energy cost of the twist-bend molecular distortion, the system transforms to the SmC. A further likely consequence of the continuous change in helical pitch on cooling is a breakup into smaller domains, resulting in loss of the initial high orientational order of the helical axis, and thus also in poor alignment in the SmC.

At the high-$T$ end, as $\phi$ decreases and the dimer straightens further, the helix unwinds at the $N_{tb}$-N transition as longitudinal motion no longer requires the molecules to follow the helical path. The drop in $\phi$ at the transition is indicated by the reduction in azimuthal spread of the WAXS arcs in Figure 3e,f.

In summary, using a specially synthesized selenium-containing compound, we have presented the first RXS experiment on the $N_{tb}$ phase using hard X-rays, which allowed magnetic sample alignment. Resonant Bragg reflection at Se absorption edge revealed a temperature dependent helical pitch of 9–12 nm. Of the two orientational order parameters required for the $N_{tb}$ phase, we find that of the helical axis, measured here for the first time, to be reaching an unprecedented value of $S_0$ = 0.99-1.00. Simultaneous recording of the resonant and the higher $q$ non-resonant scattering, impossible with carbon RXS, and the knowledge that the non-resonant features come from perfectly aligned helices allowed us to determine the tilt angle of the mesogens relative to the helical axis, as well as the bend angle of the dimers. We thus obtain a modified molecular level model of the $N_{tb}$ phase. According to it the mesogen arms of the dimer are closer to parallel than in the minimum energy conformation, and are twisted about the spacer axis. Furthermore we find that, while the segregation of the aromatic and aliphatic moieties in layers is only local, the coherently diffracting heliconical domains size is of the order of 0.1 μm.

We acknowledge funding from the joint NSF-EPSRC PIRE program ''RENEW'' (EP/K034308), the Leverhulme Trust (RPG-2012- 804) and the EPSRC (EP/M015726) and (EP/J004480) for ZA and CW and the EPSRC NMSF, Swansea for high resolution mass spectra. GU is grateful for the award of State Specially Recruited Expert from the Government of China. For help with the synchrotron experiments we thank Prof. N. Terrill at I22, Diamond Light Source, and Drs. O. Bikondoa, L. Bouchenoire, S. Brown, P. Thompson and D. Wermeille of the XMaS beamline BM28 at ESRF.

# Molecular organization in the twist-bend nematic phase by resonant X-ray scattering at the Se K-edge and by SAXS, WAXS and GIXRD

W. D. Stevenson[1], Z. Ahmed[2], X. B. Zeng[1,*], C. Welch[2], G. Ungar[1,3] and G. H. Mehl[2,*]

[1] Department of Materials Science and Engineering, University of Sheffield, Sheffield S1 3JD, UK
[2] Department of Chemistry, University of Hull, Hull HU6 7RX, UK
[3] Department of Physics, Zhejiang Sci-Tech University, Hangzhou 310018, China

## Supporting Information

**Table of Contents**

# 1.0 – Synthesis of DTSe

The Se-labelled compound DTSe was synthesised for resonant X-ray scattering (RXS) investigation using the four stage process shown diagrammatically in figure S1 below. The details of each step are also provided in the appropriate sub section. Spectroscopic infomration is listed subsequently.

Starting reagents and solvents were purchased from Fischer, Sigma Aldrich, Acros Organics and Fluorochem and used without further purification. Boronic acids were purchased from Kingston chemicals and were also used without further purification.

The structures after purification were confirmed by $^1$H and $^{13}$C nuclear magnetic resonance spectroscopy. The experiments were performed with a Joel JNM-ECP 400 MHz FT-NMR. The chemical shifts reported in this section are relative to tetramethylsilane used as an internal standard and coupling constants $J$ are reported in Hertz (Hz). $^1$H experiments were performed at 400 MHz, $^{13}$C at 100 MHz and $^{19}$F 376 MHz.

Low resolution electron ionisation (EI), electro-spray (ES), chemical ionisation (CI), matrix assisted laser deposition ionisation (MALDI) and high resolution mass spectrometry (HRMS) were obtained via the EPSRC National Mass Spectrometry Service Centre at Swansea University, Wales.

The purity of the final compounds was confirmed by high performance liquid chromatography. The HPLC setup consisted of Gilson 321 pump, Agilent/HP1100 detector with a Phenomex LUNA 18(2) reverse phase C18 column. The column dimensions are 250 mm x 4.6 mm, 5 µm particles and 100 Å pore size.

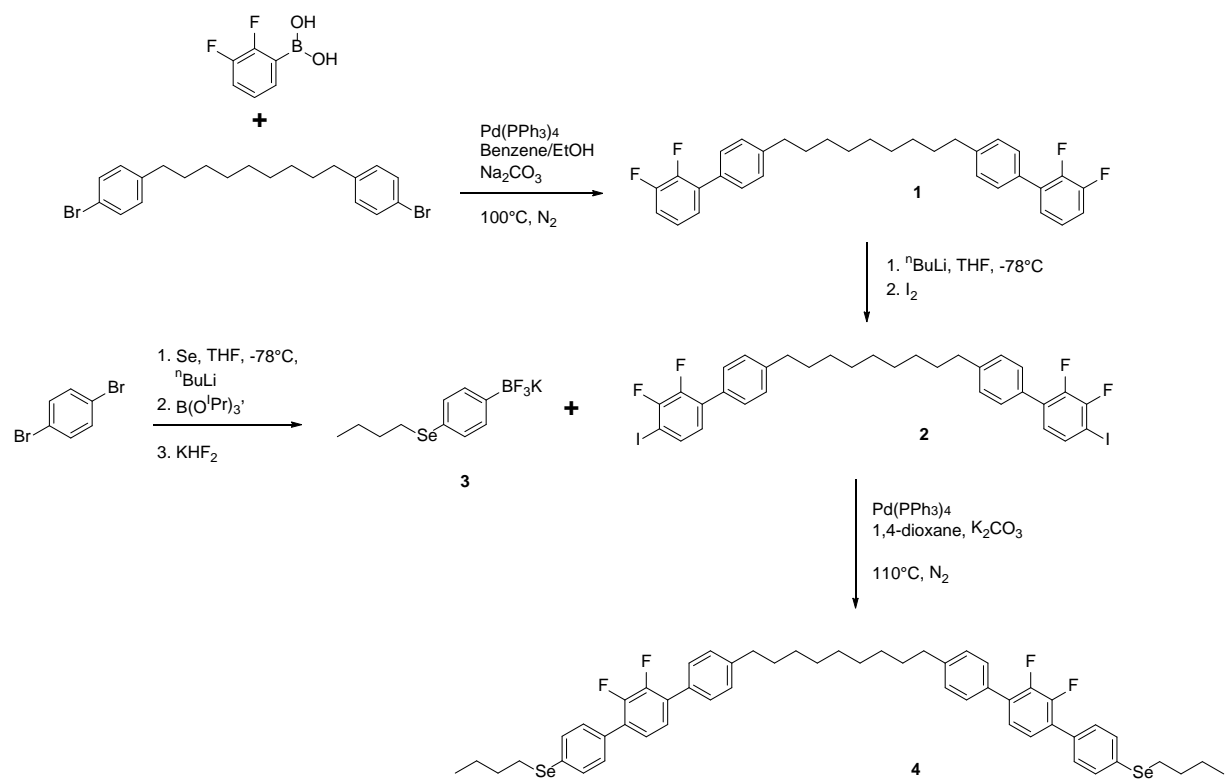

Figure S1 – The Four step Synthesis of the Se-labelled compound DTSe.

**Compound 1: 1,9-bis(2',3'-difluoro-[1,1'-biphenyl]-4-yl)nonane**

A stirred solution of 1,9-bis(4-bromophenyl)nonane (0.82g, 1.9mmol) and Pd(PPh$_3$)$_4$ (79mg, 0.07mmol, 4mol%) in benzene (15mL) were added to a solution of sodium carbonate (2M, 15mL) under nitrogen atmosphere. Nitrogen gas was bubbled through the resulting two-phase mixture for 10 minutes and then a solution of (2,3-difluorophenyl) boronic acid (0.77g, 4.9mmol) in ethanol (15mL) was added. The reaction was heated under reflux conditions to 100°C for 15 hours. After cooling to ambient temperature, diethyl ether (150mL) and water (100mL) were added to the mixture. The separated organics were then washed with water (2x100mL) and dried over MgSO$_4$(s). After removal of the dessicant by filtration, the solvent was concentrated in vacuo and the product isolated by column chromatography (Silica gel, eluent: hexane/CH$_2$Cl$_2$, 9:1). Yield 0.79g, 84%.

δ$_H$(400MHz; CD$_2$Cl$_2$) δ 7.48-7.40 (m, 4H), 7.27 (d, $^3$J=8.4Hz, 4H), 7.23-7.06 (m, 6H), 2.64 (t, $^3$J=7.8Hz, 4H), 1.63 (q, J = 7.4 Hz, 4H), 1.43-1.25 (m, 10H)

δ$_C$(100MHz; CD$_2$Cl$_2$) δ 151.58 (dd, $^1$J(C-F) = 247 Hz, $^2$J(C-F) = 13.8 Hz), 148.41 (dd, $^1$J(C-F) = 248 Hz, $^2$J(C-F) = 13.1 Hz), 143.75 (s), 132.31 (d, J(C-F) = 2.3 Hz), 131.69 (d, J(C-F) = 10.0 Hz), 129.16 (d, J(C-F) = 3.1 Hz), 129.06 (s), 125.78-125.72 (m), 124.54 (dd, J(C-F) = 7.7 Hz, 4.6 Hz), 116.07 (d, J(C-F) = 17.7 Hz), 36.05 (s), 31.88 (s), 29.94 (s), 29.91 (s), 29.76 (s)

δ$_F$(376MHz; CD$_2$Cl$_2$) -139.09 (dt, $^3$J(F-F) = 20.8 Hz, $^4$J(F-H) = 6.9 Hz), -144.75 (dd, $^3$J(F-F) = 20.8 Hz, $^4$J(F-H) = 4.6 Hz)

**Compound 2: 1,9-bis(2',3'-difluoro-4'-iodo-[1,1'-biphenyl]-4-yl)nonane.**

Compound (**1**) (0.67g, 1.3 mmol) was dissolved in anhydrous THF (15 mL) under a dry N$_2$ atmosphere and the solution was cooled to -78°C. $^n$BuLi (1.6M, 1.83 mL, 2.9 mmol) was added dropwise and (after the addition) the reaction was left to stir for 1 hour at -78°C. A solution of iodine (0.74g, 2.9 mmol) in anhydrous THF (15mL) was then added. The reaction was stirred for 16hrs, slowly rising to ambient temperature. Water (100mL) and diethyl ether (100mL) were added to separate the organics, which were then washed successively with brine (50mL) and Na$_2$S$_2$O$_3$(aq) (saturated solution, 100 mL). The organic layer was separated and dried over MgSO$_4$(s). After removal of the dessicant by filtration the solvents were concentrated to leave a yellow residue which quickly solidified. The pure product was obtained by recrystallization from ethanol.

δ$_H$(400MHz; CD$_2$Cl$_2$) 7.54 (ddd, $^3$J = 8.4 Hz, $^4$J(H-F) = 5.7 Hz, $^5$J(H-F) = 2.0 Hz, 2H), 7.42 (dd, $^3$J = 8.3 Hz, $^4$J =1.8 Hz, 4H), 7.26 (d, $^3$J = 8.4 Hz, 4H), 7.03-6.96 (m, 2H), 2.63 (t, J = 7.8 Hz, 4H), 1.69-1.57 (m, 4H), 1.39-1.27 (m, 10H)

δ$_C$(100MHz; CD$_2$Cl$_2$) 151.47 (dd, $^1$J(C-F) = 244.5 Hz, $^2$J(C-F) = 14.6 Hz), 147.84 (dd, $^1$J(C-F) = 252.9 Hz, $^2$J(C-F) = 14.6 Hz), 144.12 (s), 133.72 (d, J(C-F) = 3.8 Hz), 131.97 (d, J(C-F) = 10.8 Hz), 131.53 (br s), 129.12 (s), 128.96 (d, J(C-F) = 3.1 Hz), 126.89 (m), 80.23 (d, $^2$J(C-F) = 23.1 Hz), 35.69 (s), 31.77 (s), 29.83 (s), 29.80 (s), 29.65 (s)

δ$_F$(376MHz; CD$_2$Cl$_2$) -117.92 (dd, $^3$J(F-F) = 20.8 Hz, $^4$J(F-H) = 4.6 Hz), -140.06 (dd, $^3$J(F-F) = 20.8 Hz, $^4$J(F-H) = 6.9 Hz)

MS (EI) *m/z* 756.0 (M)$^+$
HRMS : calculated for C$_{33}$H$_{30}$F$_4$I$_2$ : 756.0373, found 756.0373

**Compound 3: Potassium 4-(Butylselanyl)phenyltrifluoroborate.**

A solution of 1,4-dibromobenzene (2.36g, 10 mmol) and selenium powder (0.79 g, 10 mmol) in anhydrous THF (60 mL) was cooled to -78°C in a nitrogen atmosphere. $^n$BuLi (2.2 M in cyclohexane, 4.5 mL, 10 mmol) was added dropwise and the reaction was warmed to 0°C over a period of 30 minutes. After the reaction mixture was clear, triisopropyl borate (1.88 g, 2.31 mL, 10 mmol) was added (at 0°C) and the reaction mixture was re-cooled to -78°C. ⁿBuLi (2.2 M in cyclohexane, 4.5 mL, 10 mmol) was then slowly added. After 10 minutes at -78°C the temperature was raised to -10°C and the reaction was stirred at this temperature for 40 minutes. The reaction mixture was quenched with KHF$_2$(aq) (1N, 25 mL, 25 mmol) and then warmed to room temperature. After stirring for 15 minutes, the suspension was concentrated and the residual solid was dissolved in dry acetone (80 mL). The insoluble salts were filtered off and the resulting organic solution was concentrated to leave a solid. This was dissolved in acetone (70 mL) and diethyl ether (70 mL) and precipitated to produce an off-white solid, 1.64g, 51%

$\delta_H$(400MHz; (CD$_3$)$_2$CO)  7.41 (d, $^3J$ = 7.7 Hz, 2H), 7.26 (d, $^3J$ = 7.6 Hz, 2H), 2.82 (t, $^3J$ = 7.3 Hz, 2H), 1.62 (m, 2H), 1.40 (m, 2H), 0.87 (t, $^3J$ = 7.3 Hz, 3H)

$\delta_C$(100MHz; (CD$_3$)$_2$CO) 133.42, 132.07, 126.78, 33.17, 27.88, 23.46, 13.82

$\delta_F$(376MHz; (CD$_3$)$_2$CO) -142.54

$\delta_B$(128 MHz; (CD$_3$)$_2$CO) 2.69

MS (NSI) *m/z* 274.03 (M-K)⁻
HRMS : calculated for C$_{10}$H$_{13}$$^{10}$BF$_3$$^{74}$Se : 274.0329, found 274.0335

**Compound 4: 1,9-bis(4''-(butylselanyl)-2',3'-difluoro-[1,1':4',1''-terphenyl]-4-yl)nonane.**

Pd(PPh$_3$)$_4$ (28 mg, 0.02mmol, 6mol%) and K$_2$CO$_3$ (0.33 g, 2.4 mmol) were added to a mixture of (**2**) (0.3g, 0.4 mmol) and (**3**) (0.25g, 0.8 mmol) dissolved in 1,4-dioxane (32 mL) and water (8 mL). Nitrogen gas was bubbled through the mixture for 10 minutes and the reaction was subsequently heated to 110°C for 15 hours under an N$_2$ atmosphere. After cooling to ambient temperature, the organics were separated by adding diethyl ether (150mL) and water (100mL). The organics were washed with water (2x100mL) and dried over MgSO$_4$(s). After removal of the dessicant by filtration, the solvent was concentrated in vacuo and the product isolated by column chromatography (Silica gel, eluent: hexane/CH$_2$Cl$_2$, 8:2). Yield 0.25 g, 68%.

$\delta_H$(400MHz; CD$_2$Cl$_2$)  7.58-7.54 (m, 4H), 7.52-7.46 (m, 8H), 7.30 (d, $^3J$ = 8.2 Hz, 4H), 7.28-7.26 (m, 4H), 2.99 (t, $^3J$ = 7.3 Hz, 4H), 2.67 (t, $^3J$ = 7.6 Hz, 4H), 1.77-1.61 (m, 8H), 1.51-1.29 (m, 14H), 0.93 (t, $^3J$ = 7.3 Hz, 6H)

$\delta_C$(100MHz; CD$_2$Cl$_2$) 148.93 (dd, $^1J$ = 249.8Hz, $^2J$ = 15.4Hz), 148.75 (dd, $^1J$ = 249.8 Hz, $^2J$ = 15.4 Hz), 143.85 (s), 133.04 (s), 132.20 (s), 132.00 (s), 130.22 (dd, J(C-F) = 8.5 Hz, 2.3 Hz), 129.68 (s), 129.30-129.19 (m), 129.12 (s), 129.04 (br s), 125.24-125.17 (m), 124.97-124.90 (m), 36.03 (s), 32.64 (s), 31.85 (s), 29.86 (s), 29.83 (s), 29.69 (s), 27.67 (s), 23.39 (s), 13.73 (s)

$\delta_F$(376MHz; CD$_2$Cl$_2$) -143.92 (s)

MS (EI) *m/z* 928.1 (M)⁺

## 1.1. – Chemical Analysis -Spectroscopic data for compounds 1-4

**Compound 1**

**$^1$H-NMR data for compound 1**

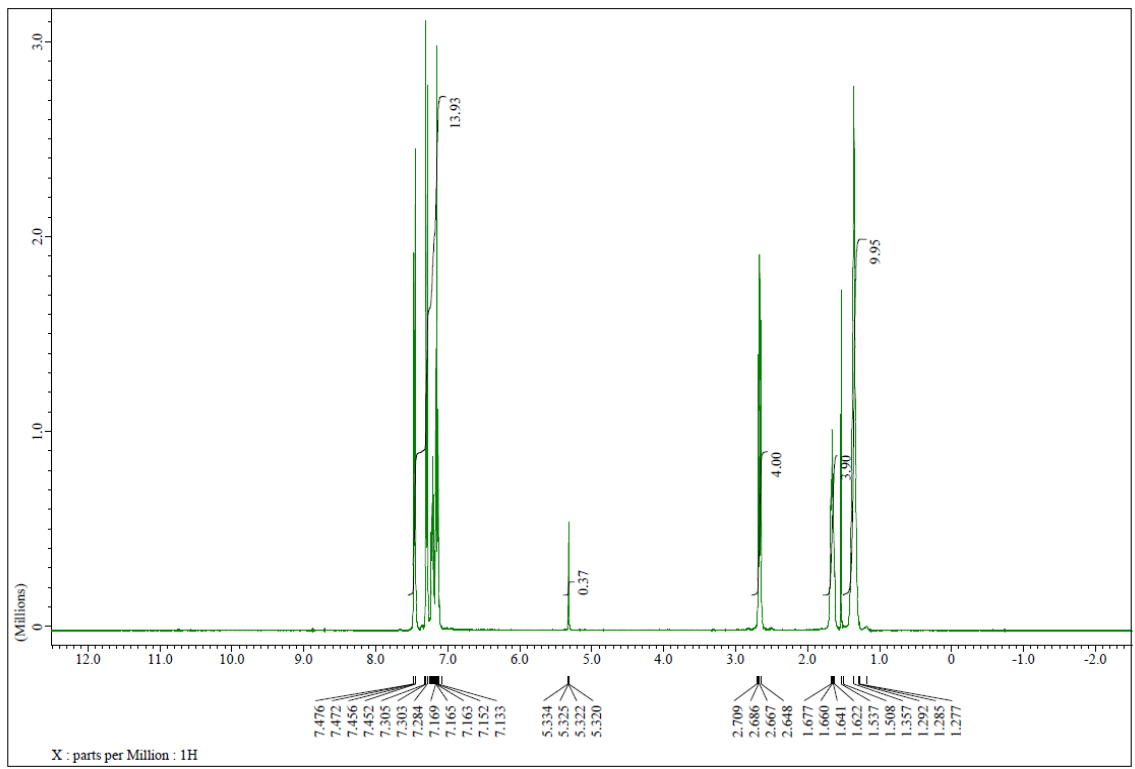

**$^{13}$C- NMR data for compound 1**

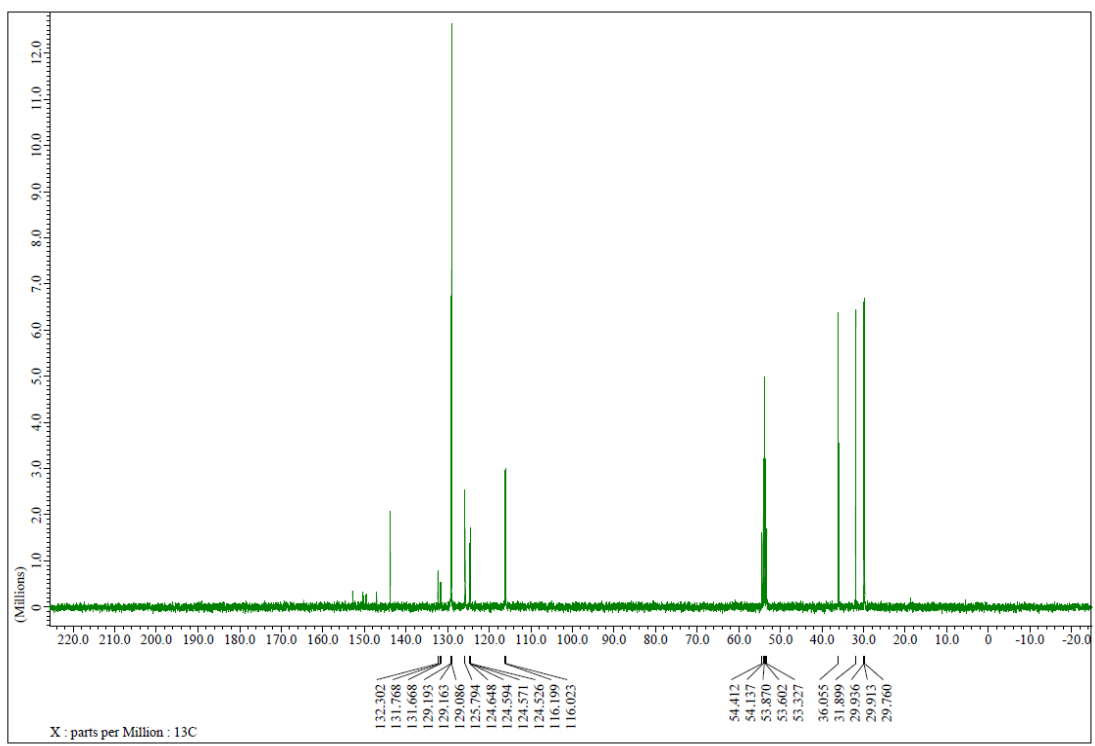

## Compound 2

### ¹H-NMR data for compound 2

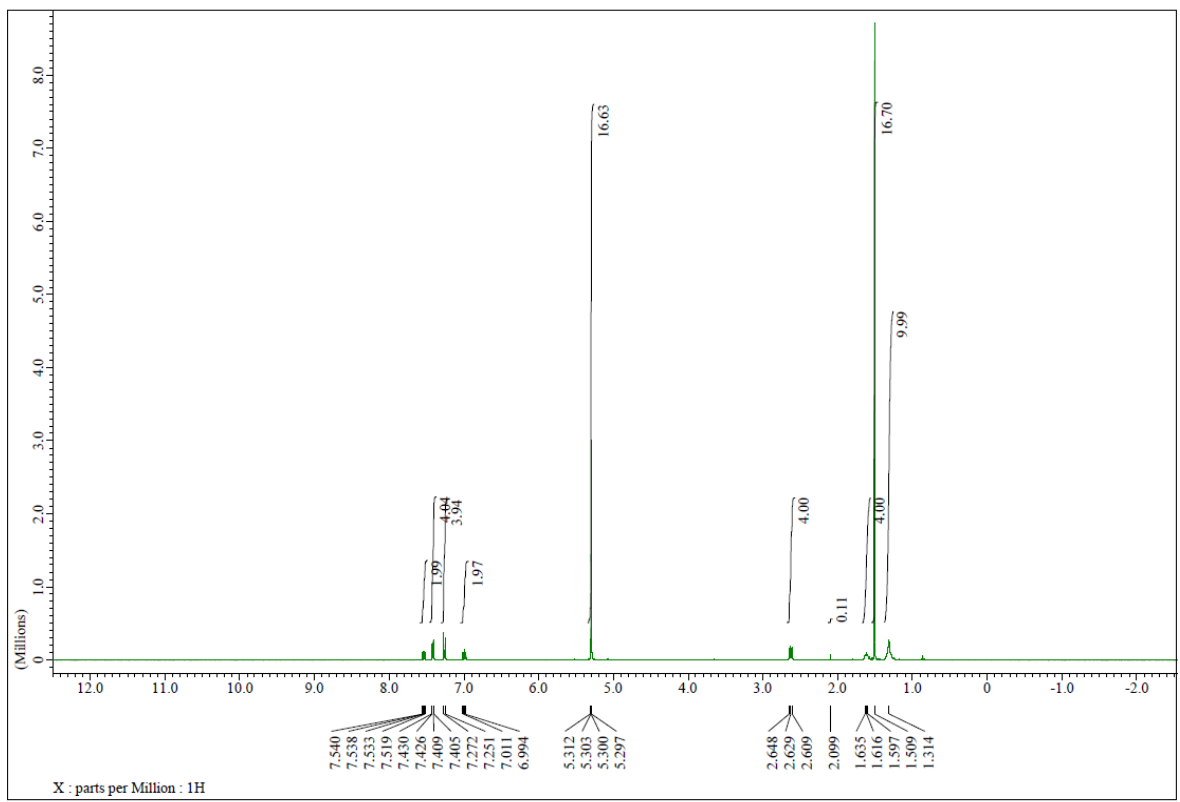

### ¹³C- NMR data for compound 2

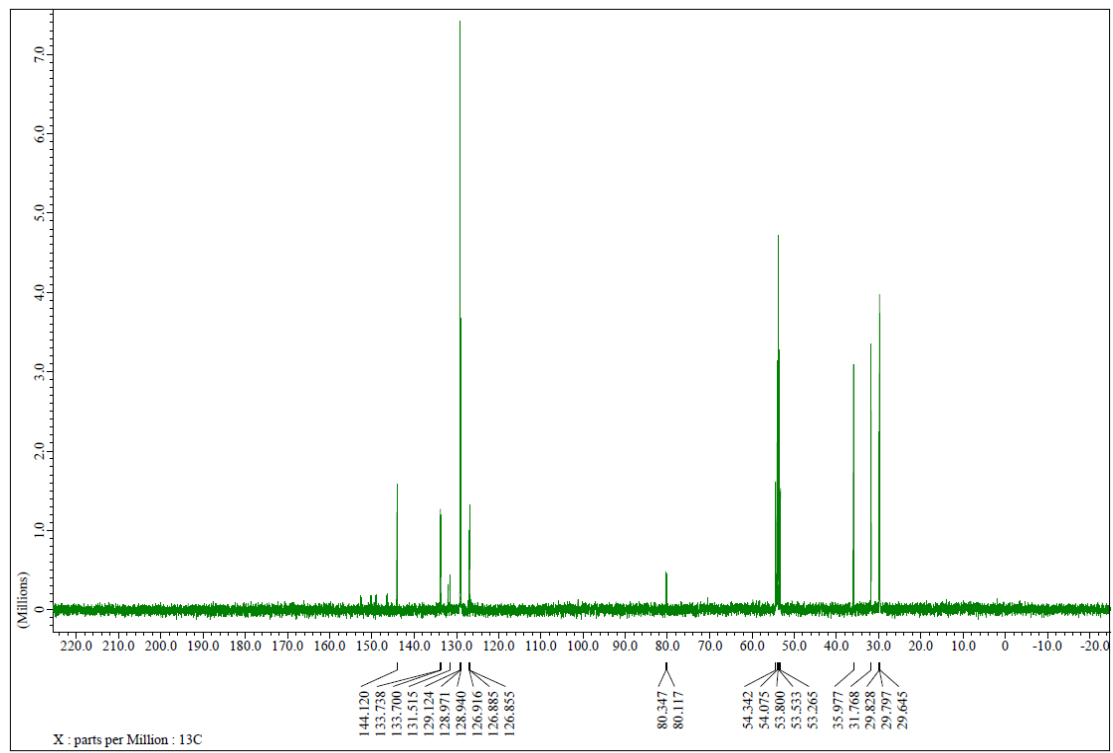

## 19F –NMR data for compound 2

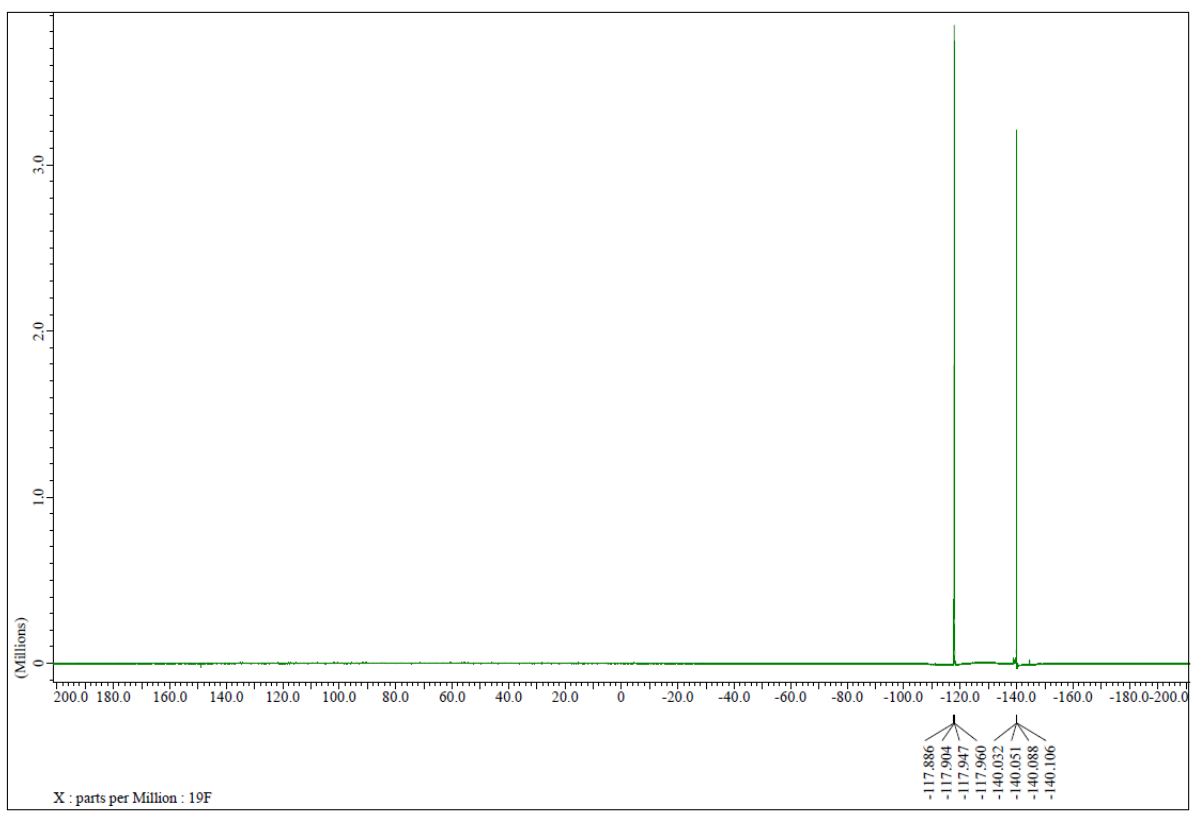

## High Resolution Mass spectroscopy data for compound 2

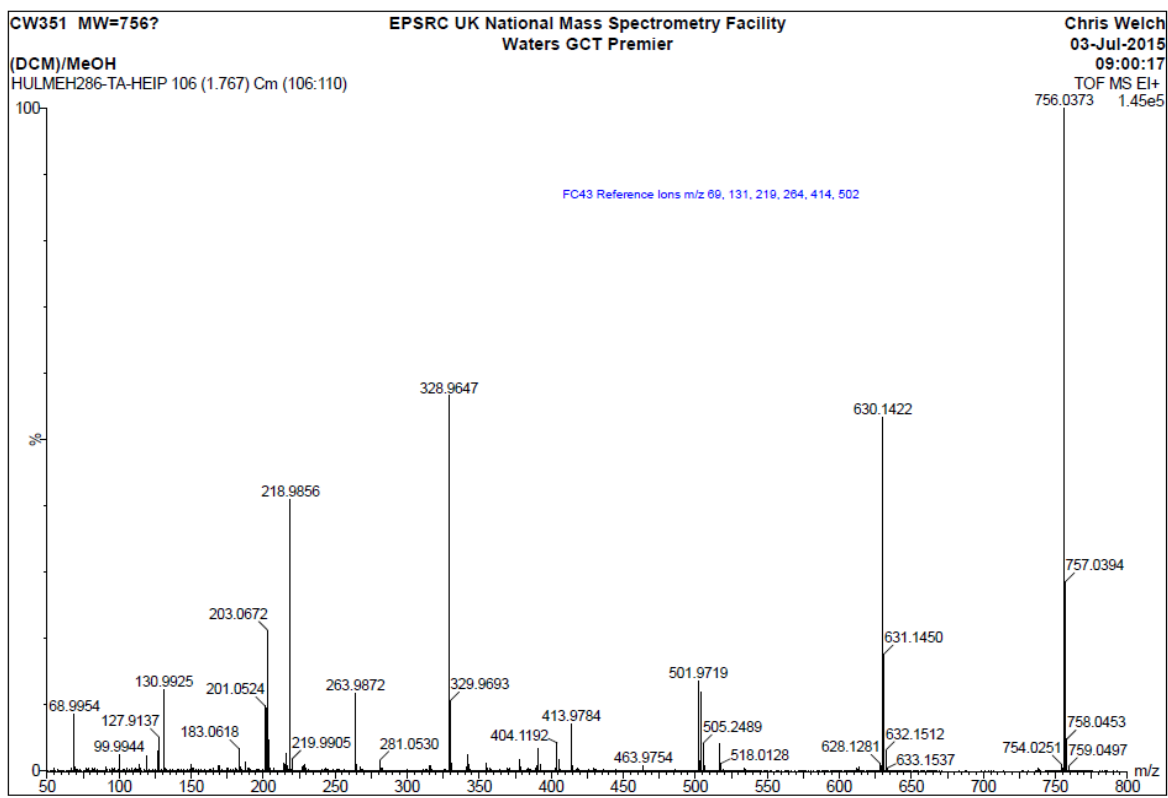

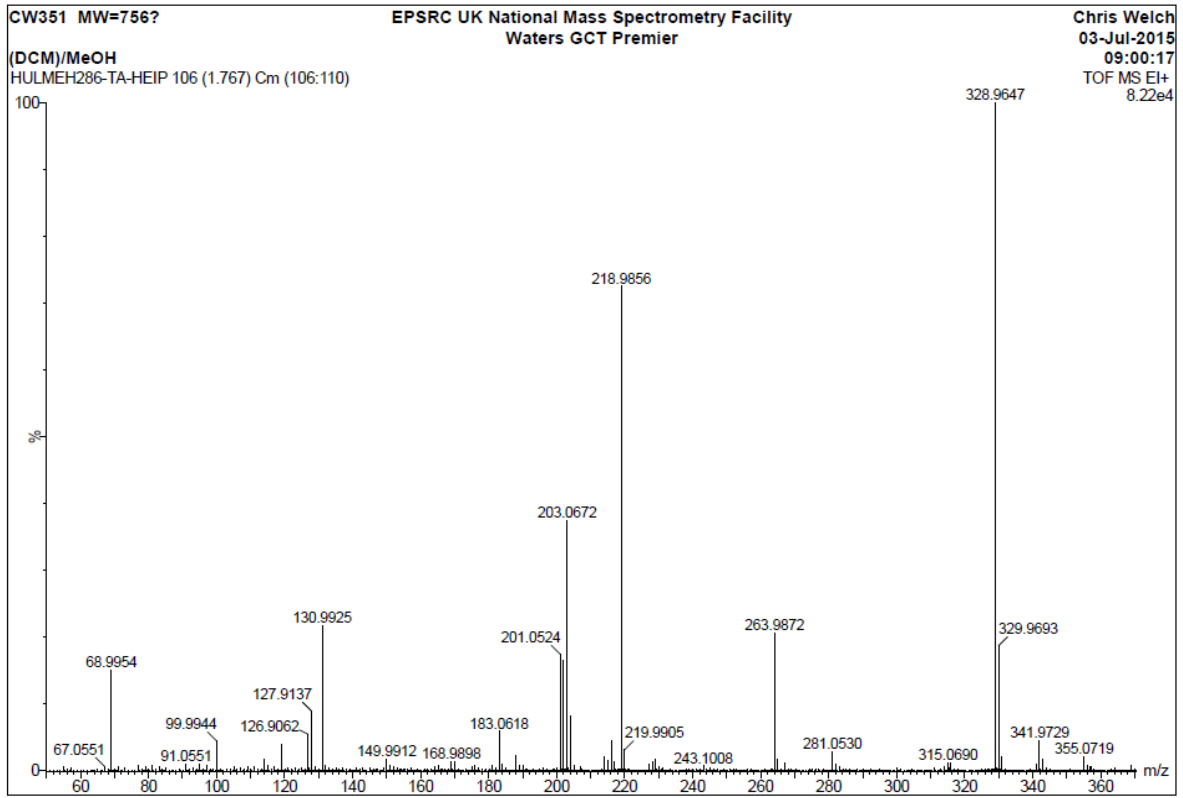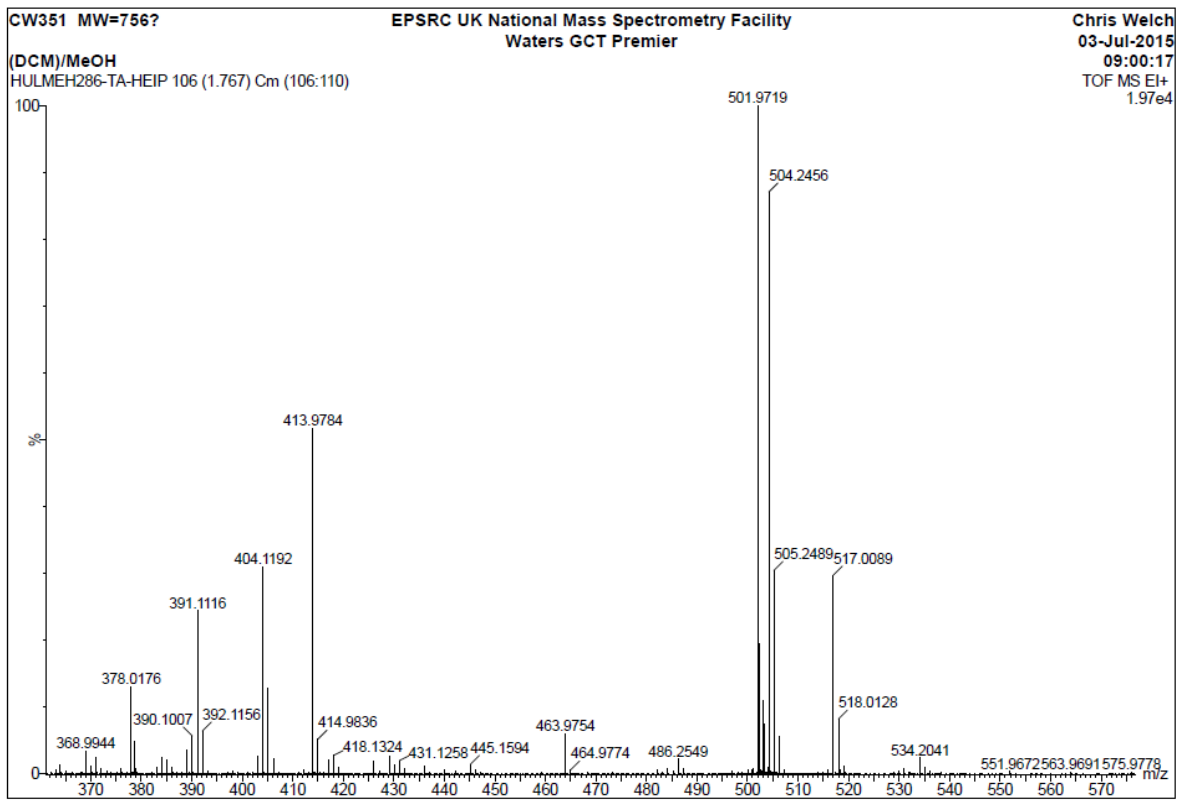

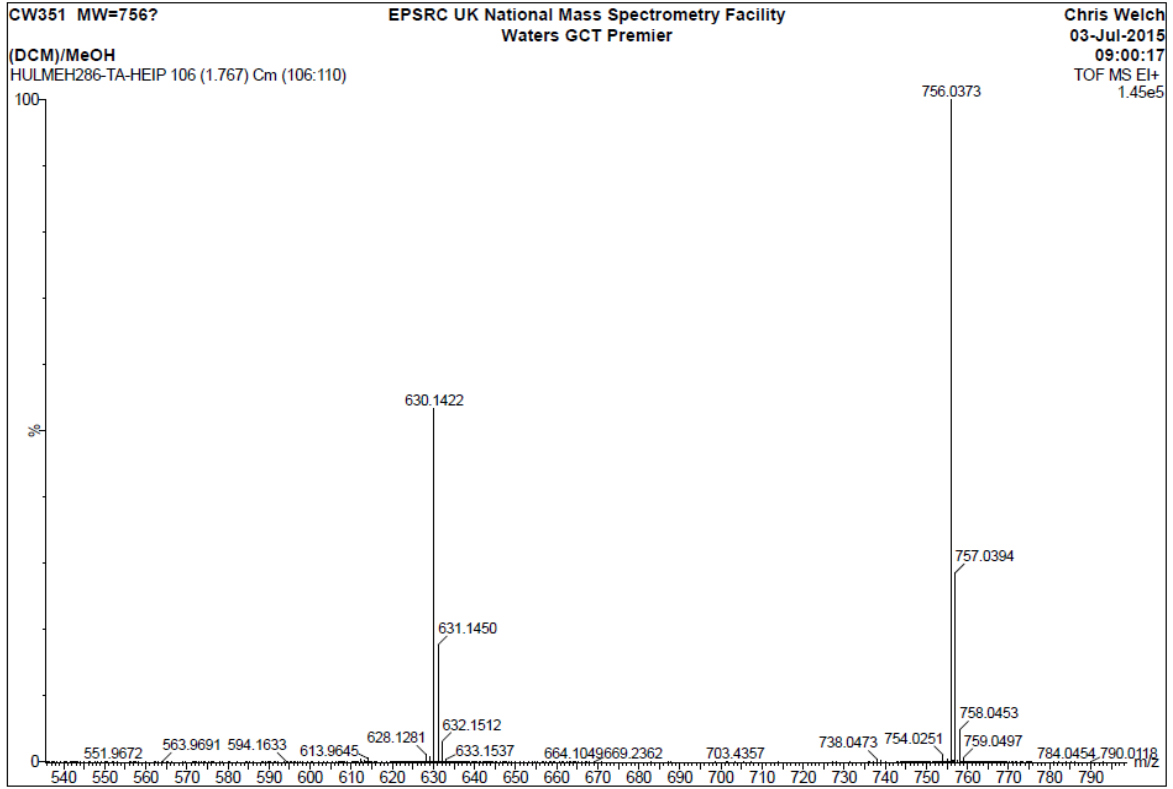

Comparison of theoretical and experimental isotope profile for compound 2

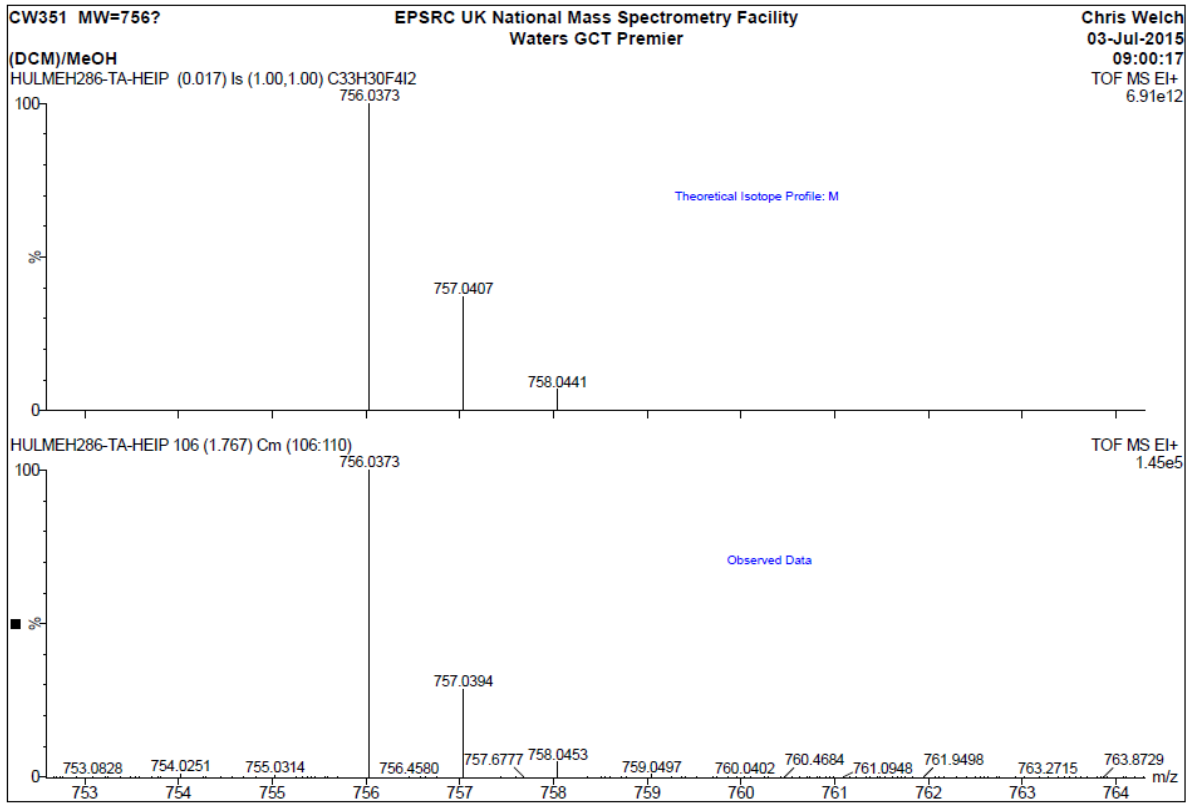

## Compound 3

### ¹H-NMR data for compound 3

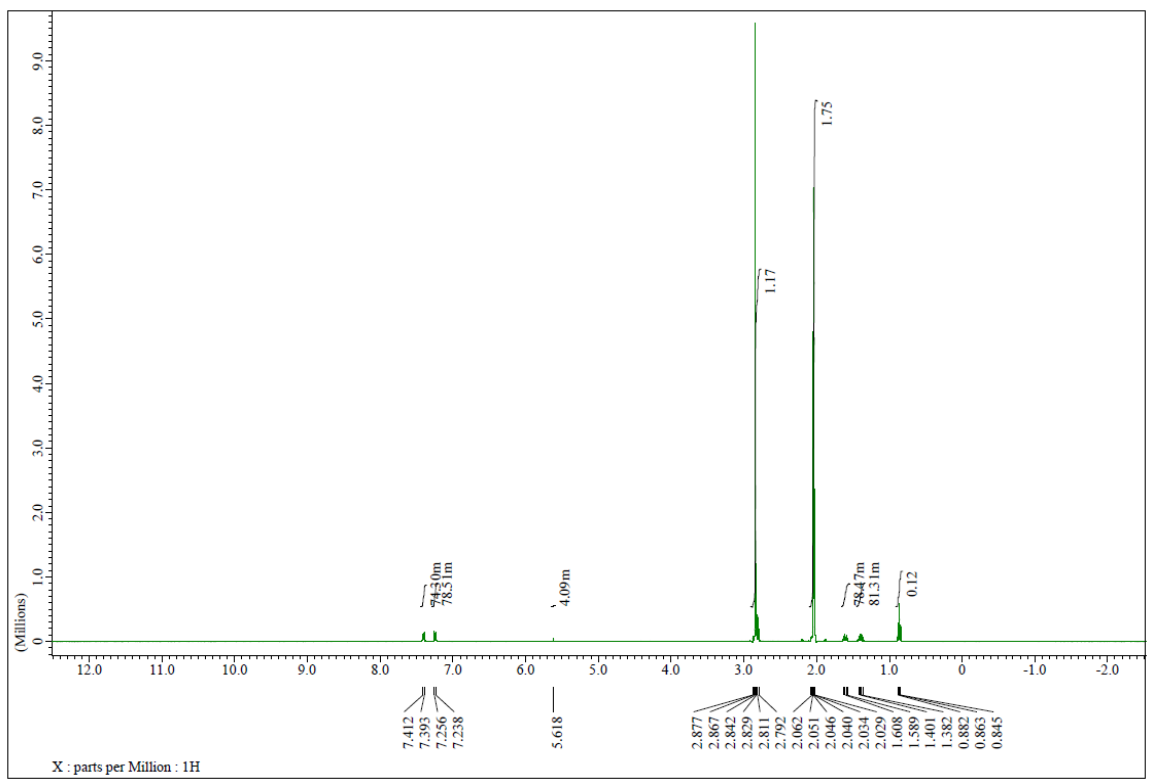

### ¹³C- NMR data for compound 3

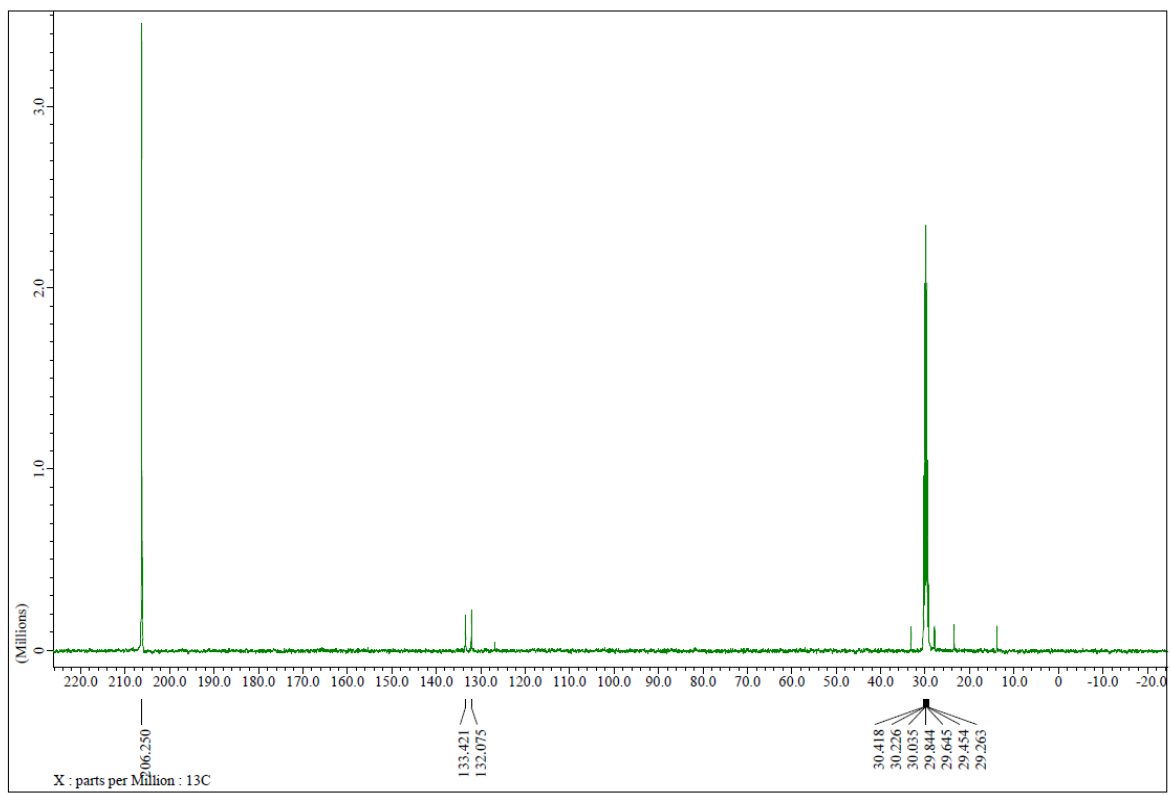

## ¹¹B- NMR data for compound 3

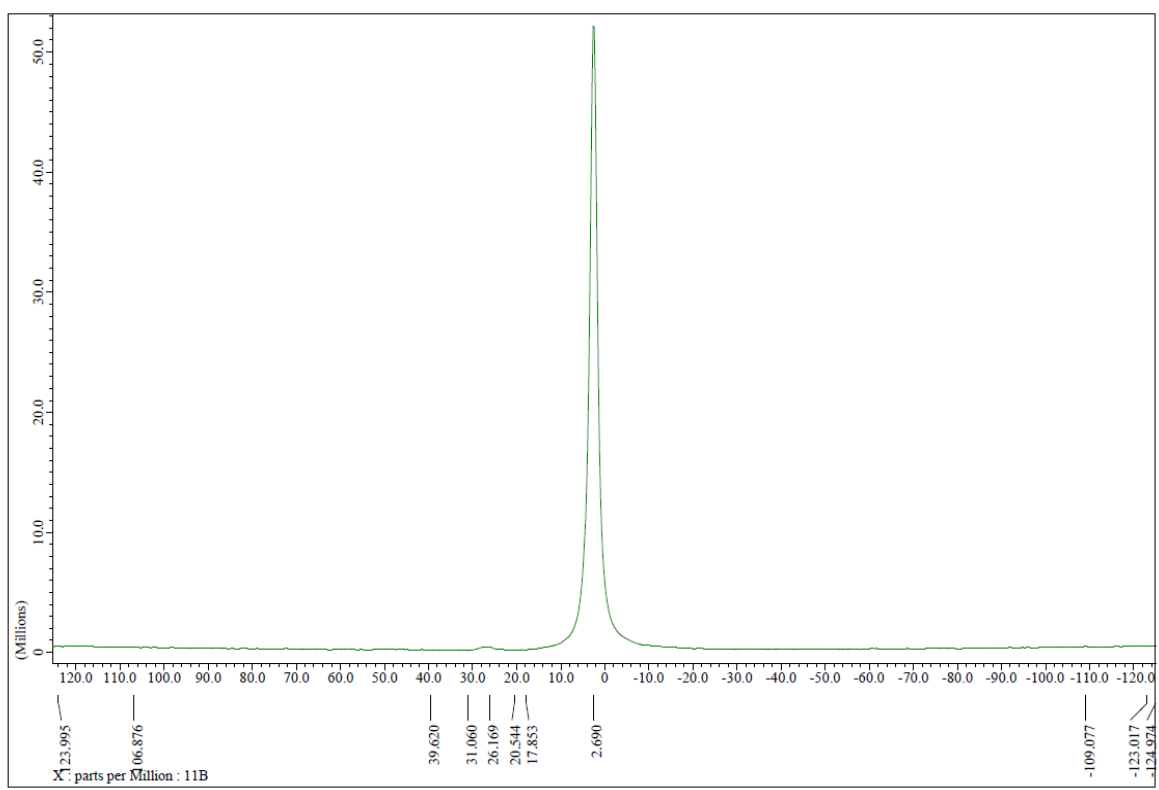

## ¹⁹F –NMR data for compound 3

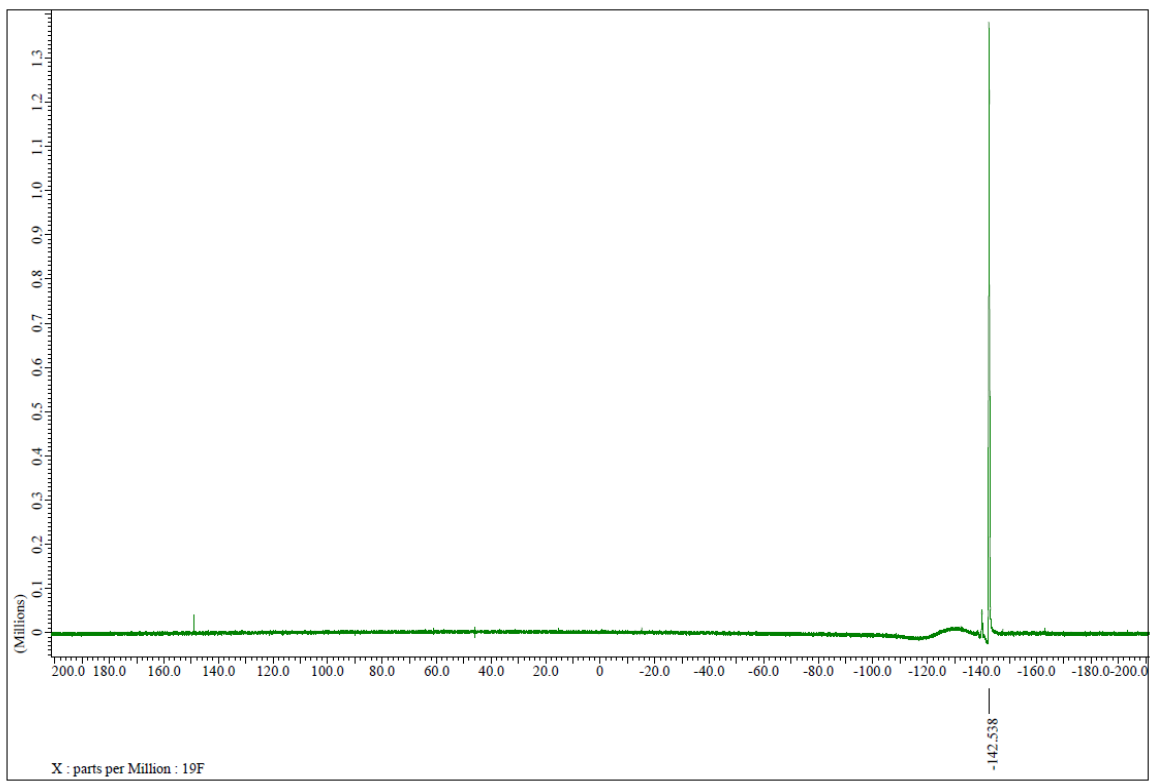

## High Resolution Mass spectroscopy data for compound 3

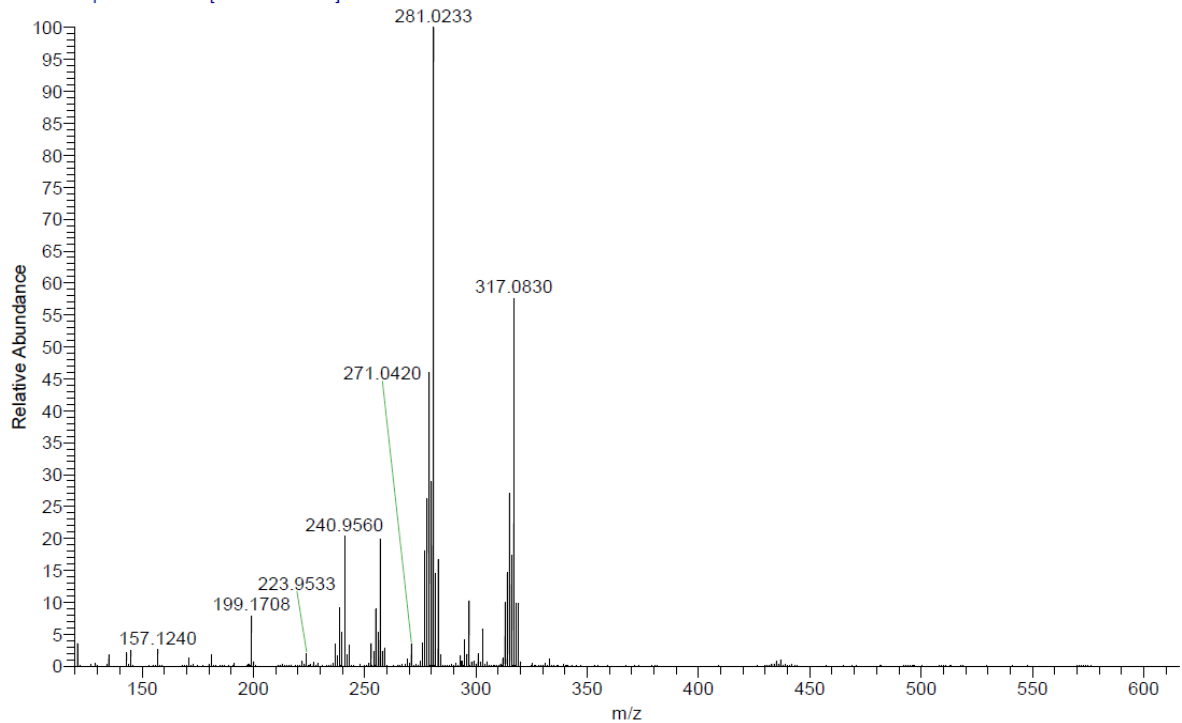

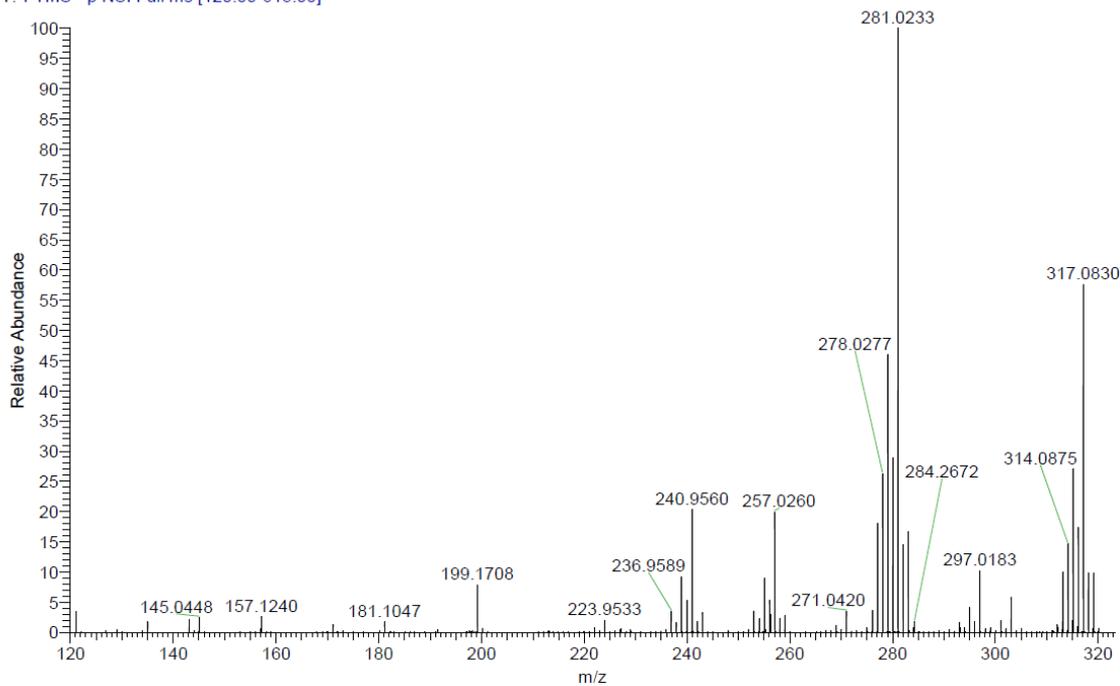

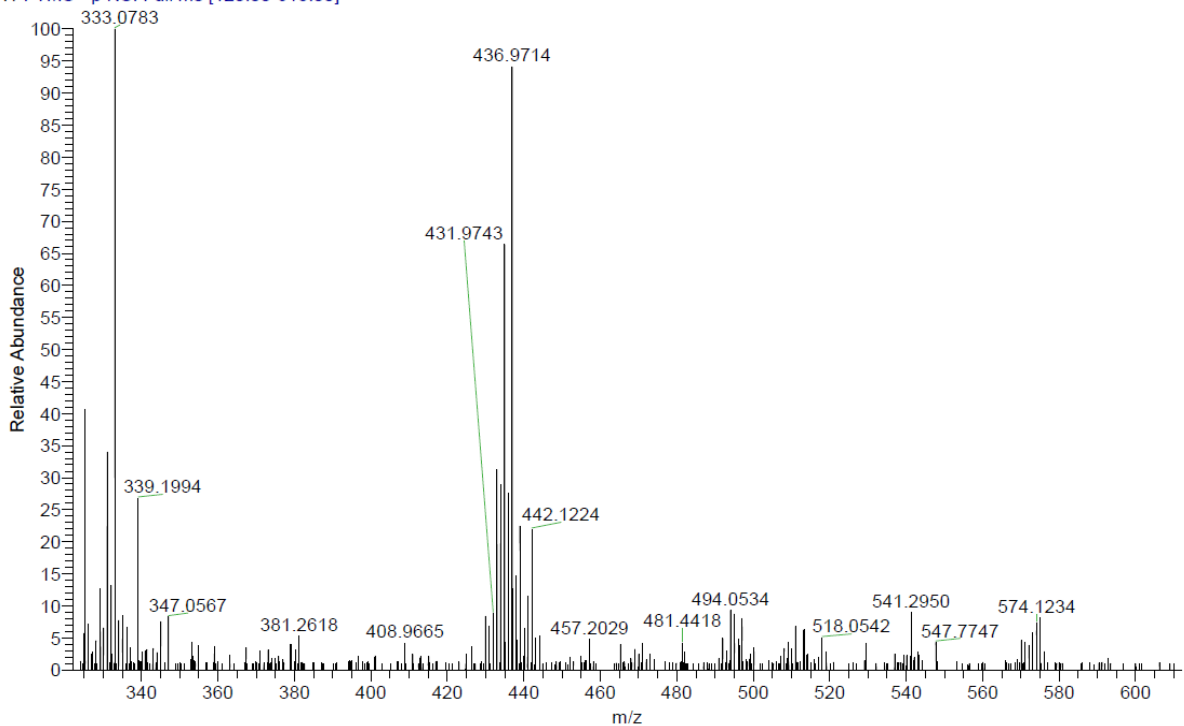

Comparison of theoretical and experimental isotope profile for compound 3

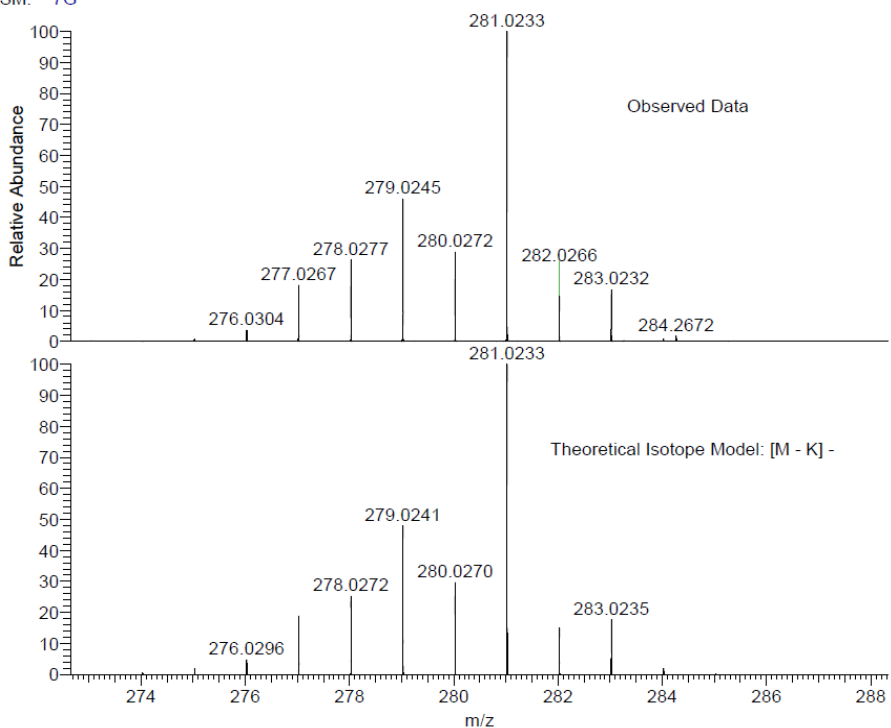

## Compound 4

### ¹H-NMR data for compound 4

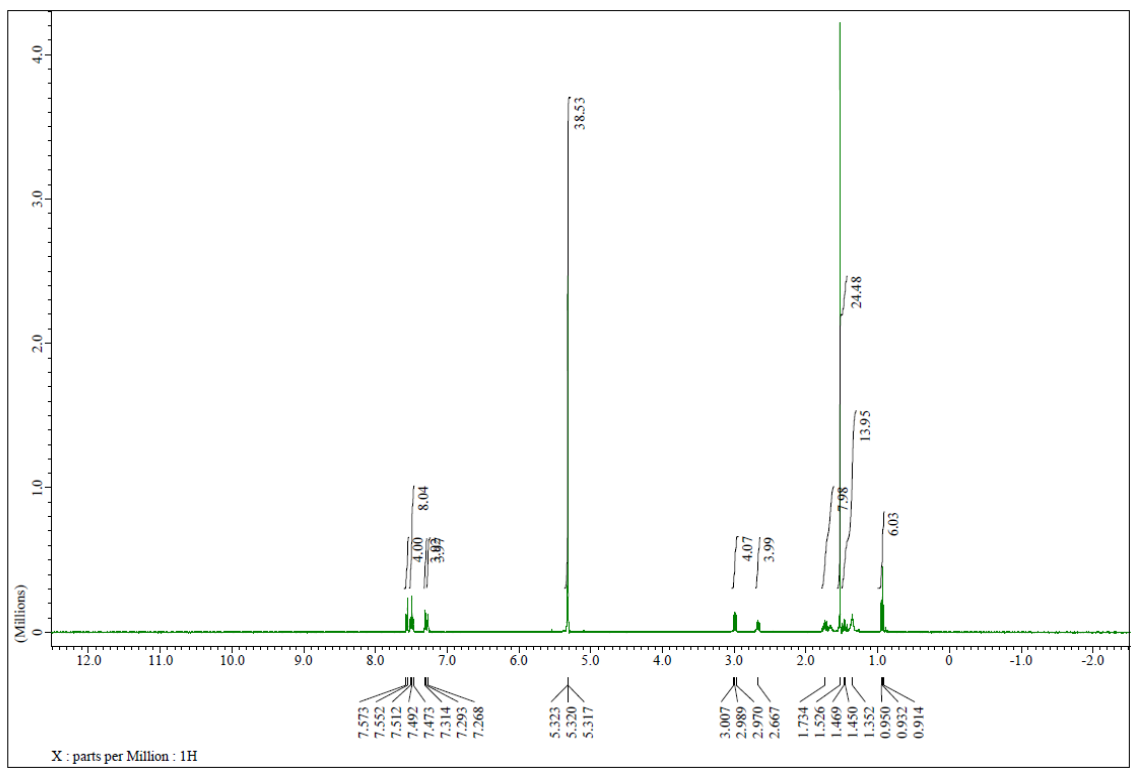

### ¹³C- NMR data for compound 4

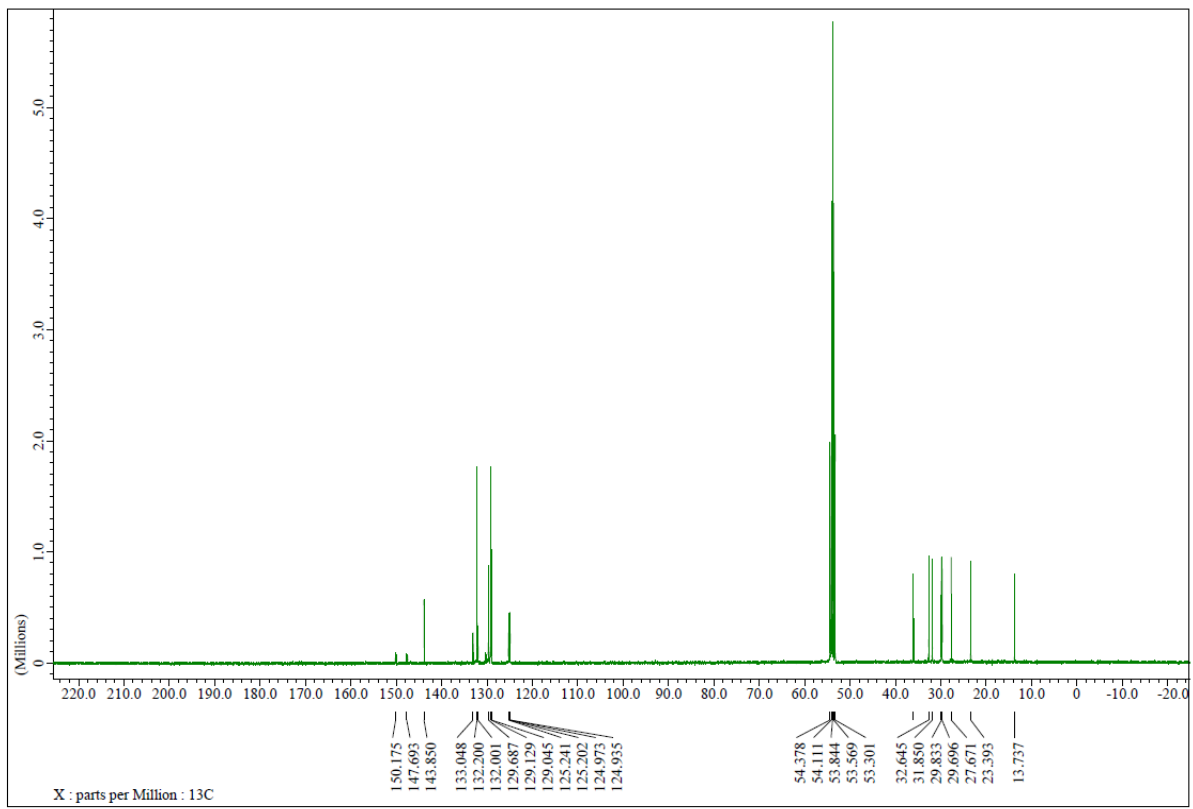

## 19F –NMR data for compound 4

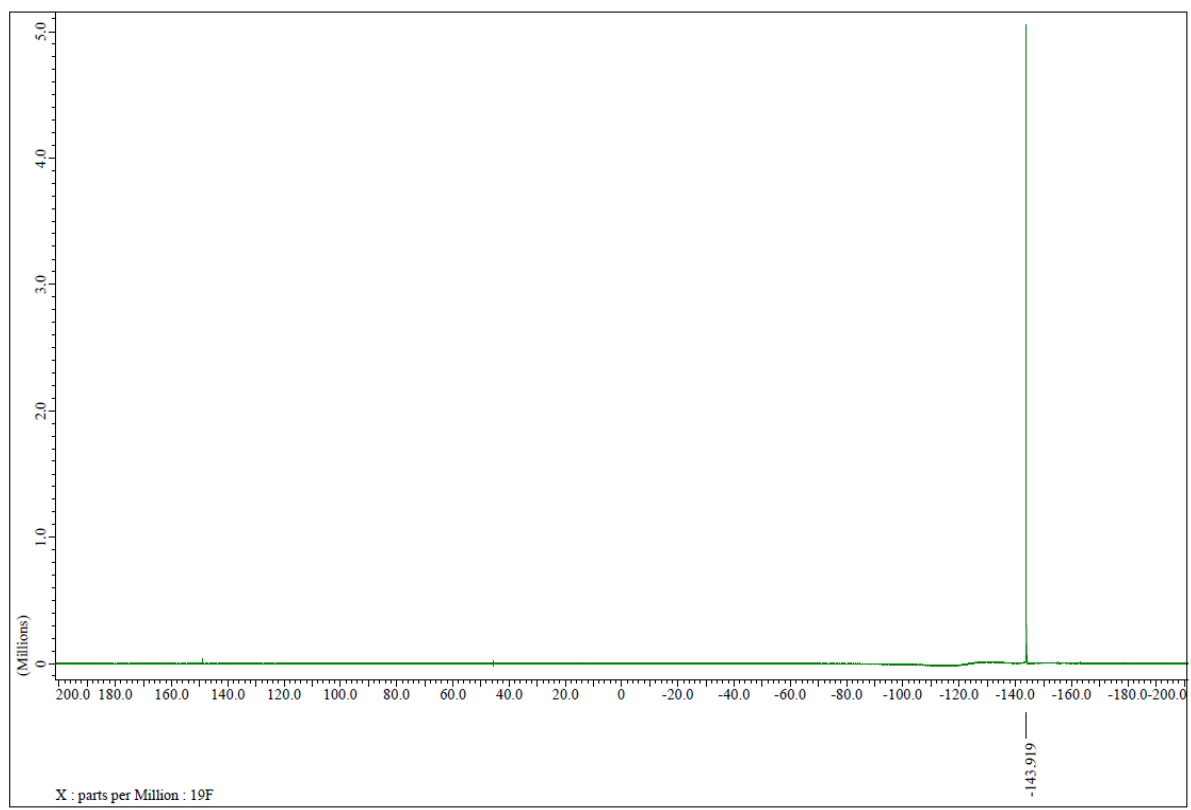

## High Resolution Mass spectroscopy data for compound 4

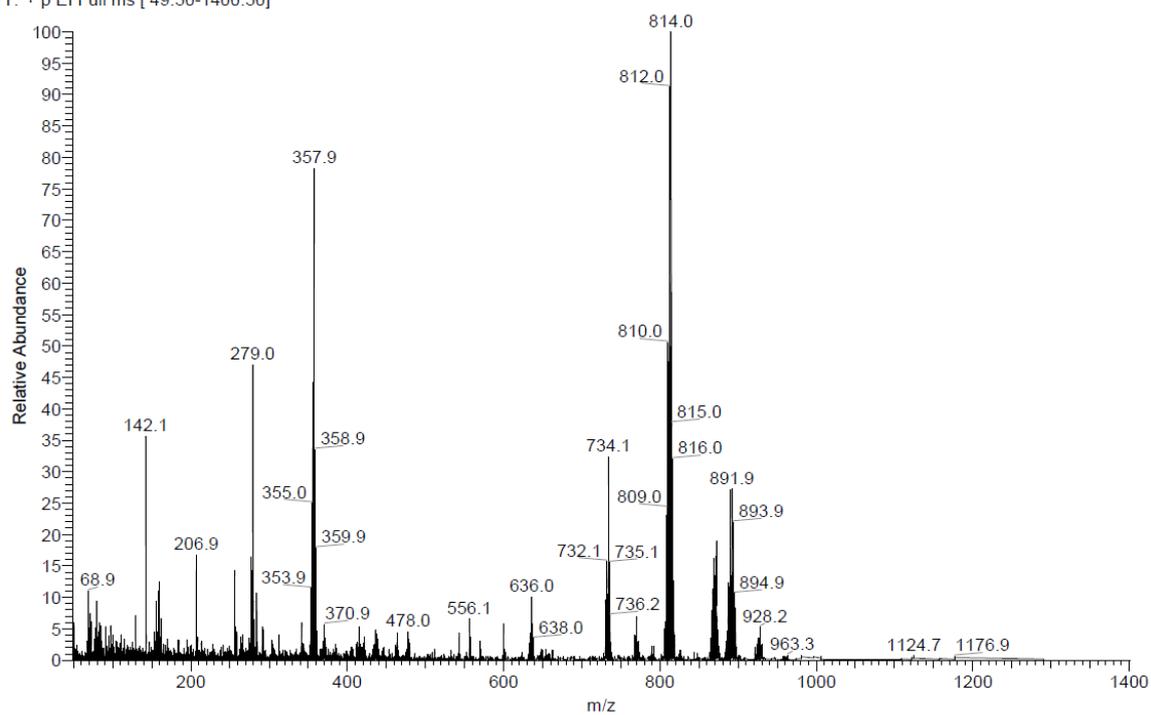

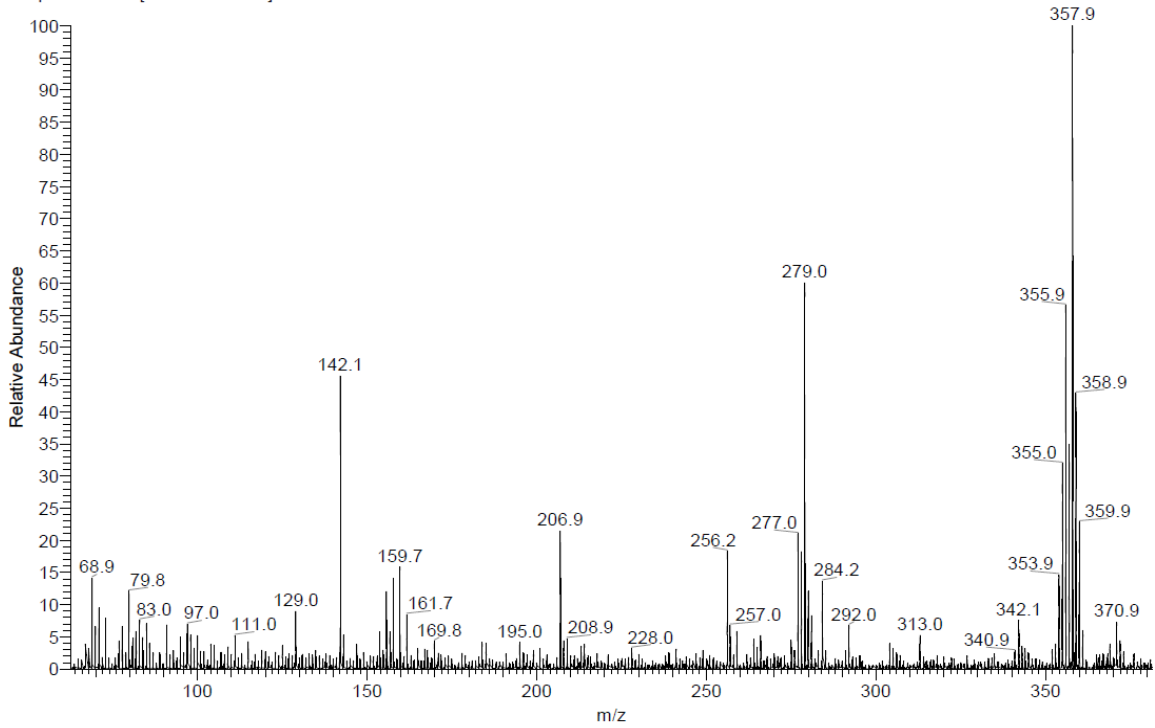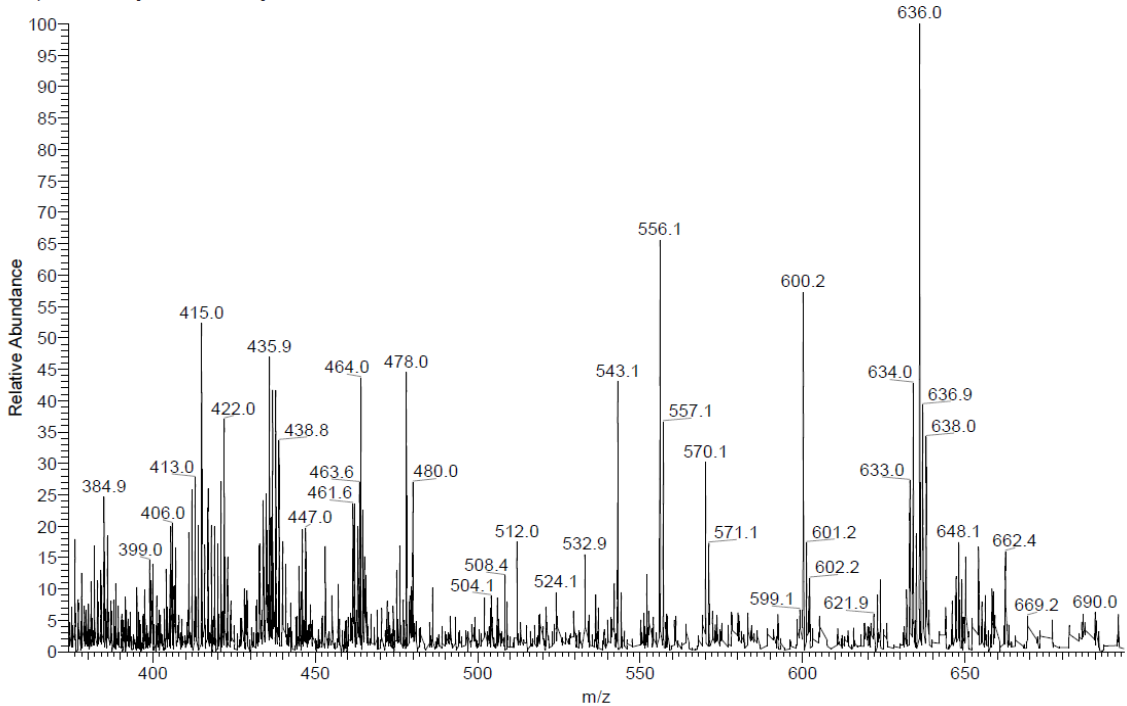

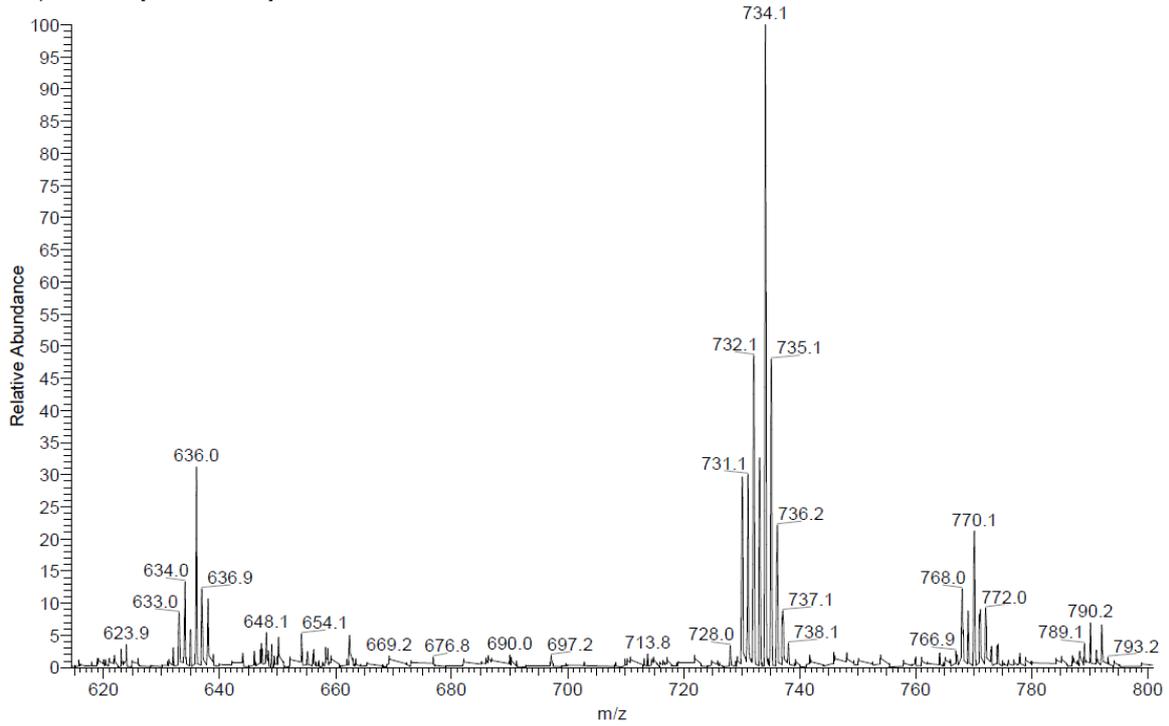
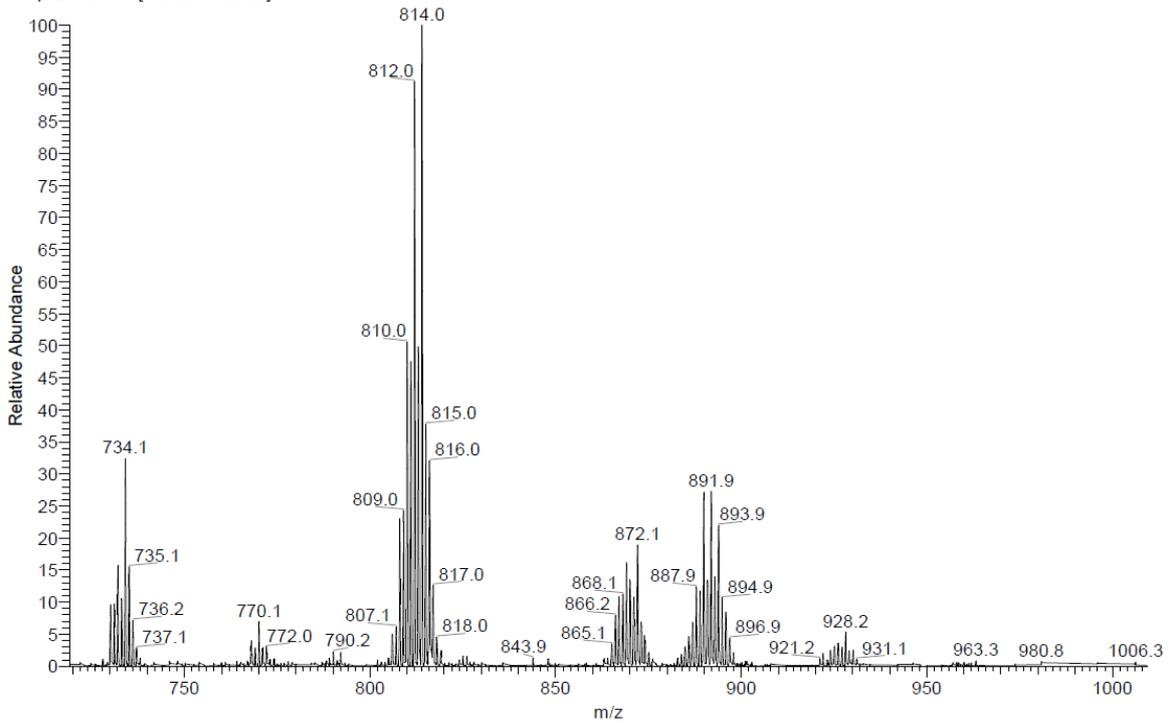

**Comparison of theoretical and experimental isotope profile for compound 4**

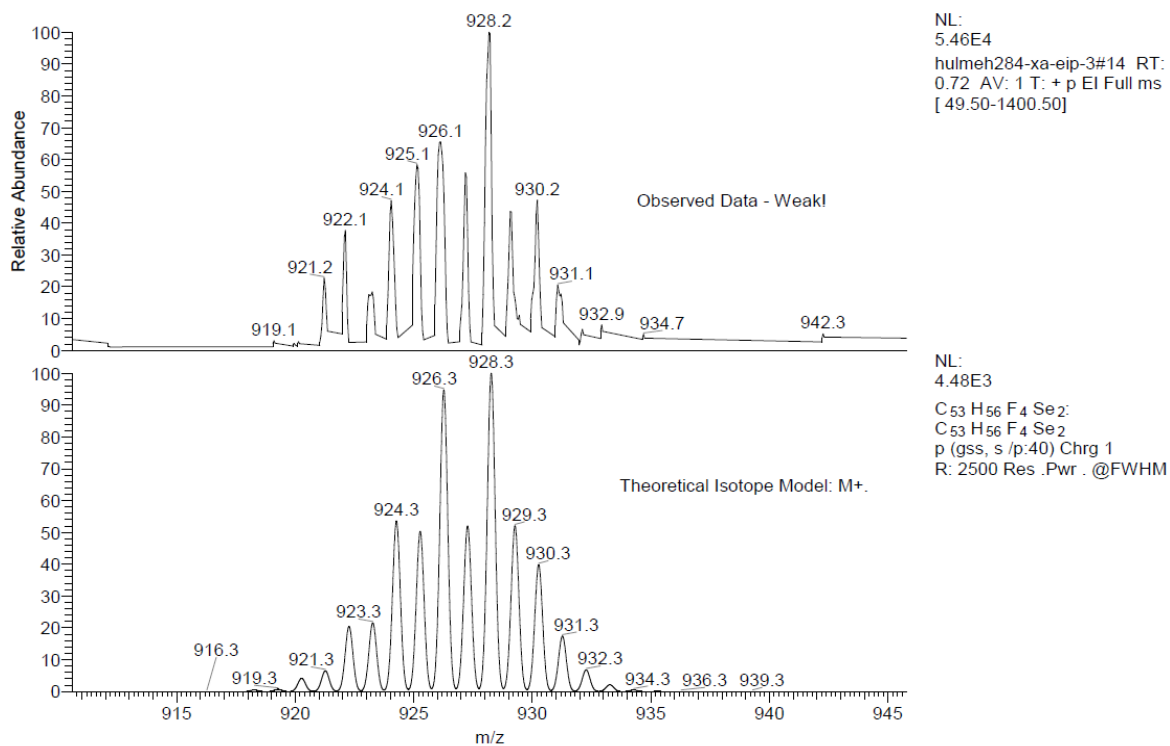

**Note:**

Accurate mass measurement is not possible, as the monoisotopic ions of the multiple Se multiisotopic envelope are too weak to measure accurately with current technology. Instead the isotope profile is used instead, see above. This is an issue for compounds which contain more than one Se-atom.

## 2.0 – Miscibility of DTC5C7 and DTSe

The miscibility of DTC5C7 and DTSe was examined under crossed polarisers using an Olympus BX50 Microscope fitted with a Mettler FP82 HT hotstage. A small amount of both samples was sandwiched between two glass plates, each 1mm in thickness. The glass cell (~1cm × 1.5cm) was heated to 160°C and compressed slightly in order to create a region of contact between the two compunds. The cell was then cooled at 10°C/min to 60°C. The key transition points (not shown in the main text) are displayed below in Figure S2.

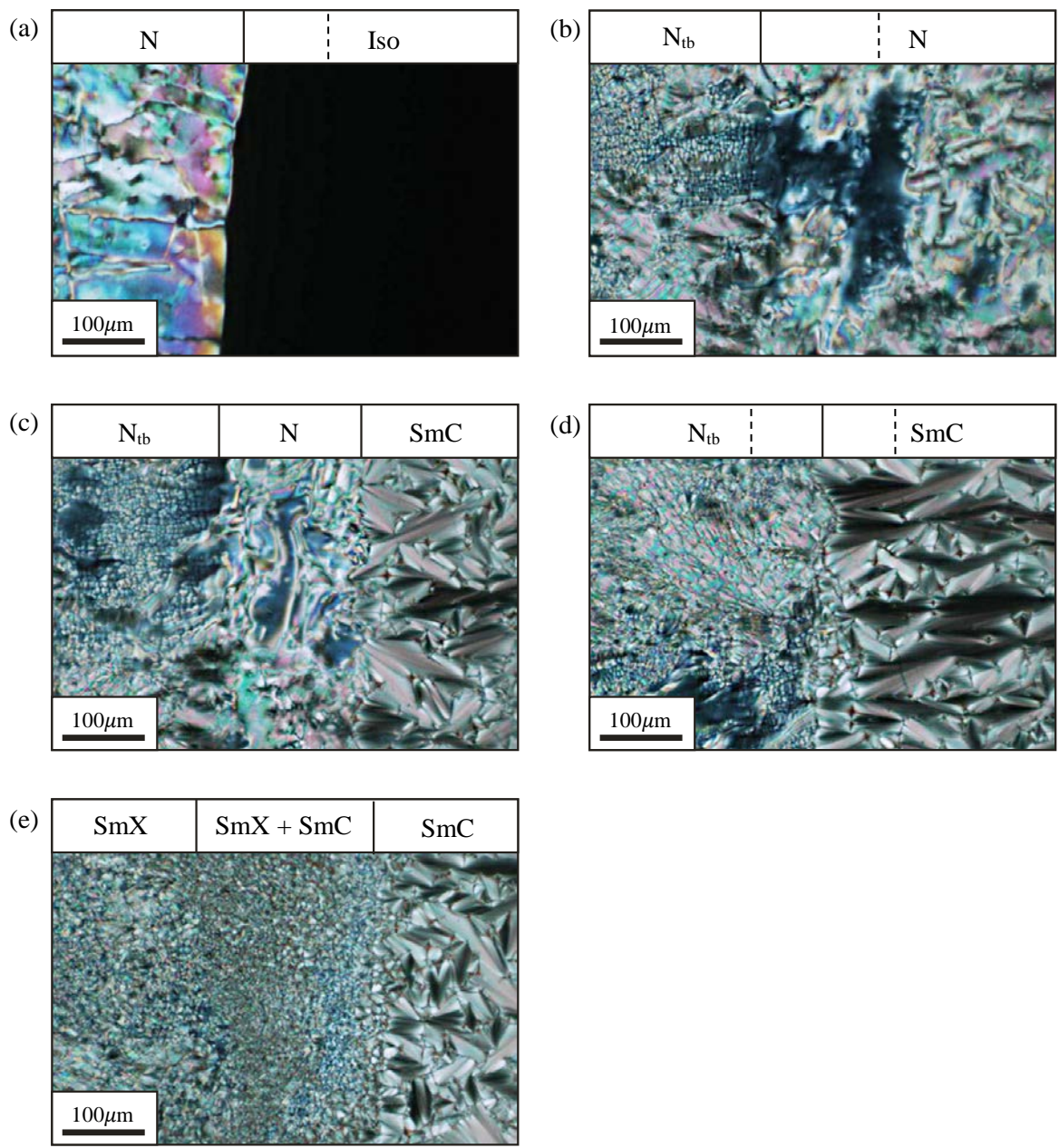

Figure S2 – The optical textures observed between crossed polarisers in a sandwich cell containing both DTC5C7 and DTSe. In all four images DTSe is on the right and DTC5C7 is on the left. The central region is a mixture of the two. Solid lines indicate phase boundaries and dashed lines indicate the approximate edge of the mixed region when it is in the same phase as one (or more) of the pure compounds. The images were taken at (a) 130°C, (b) 120°C, (c) 114°C, (d) 110°C and (e) 90°C. Iso = Isotropic, N = nematic, $N_{tb}$ = twist bend nematic, SmC = smectic C, SmX = modulated smectic. SmX is meta-stable and only observed on cooling.

## 3.0 – DSC Analysis of DTC5C7, DTSe and Mixtures

The phase behaviour of all compounds and mixtures was investigated by differential scanning calorimetry (DSC) using a Perkin Elmer Diamond DSC, fitted with a ULSP liquid nitrogen cooling unit. The scanning rate was 10°C/min in all instances. The samples were investigated on both heating and cooling. The first heating curves are displayed below in Figure (S1) and were used to construct the phase diagram displayed in Figure 1(c) of the main text. The transition temperatures are tabulated in Table S1. The mixtures are named using the notation 'Se00', where the last two digits correspond to the weight percent of DTSe.

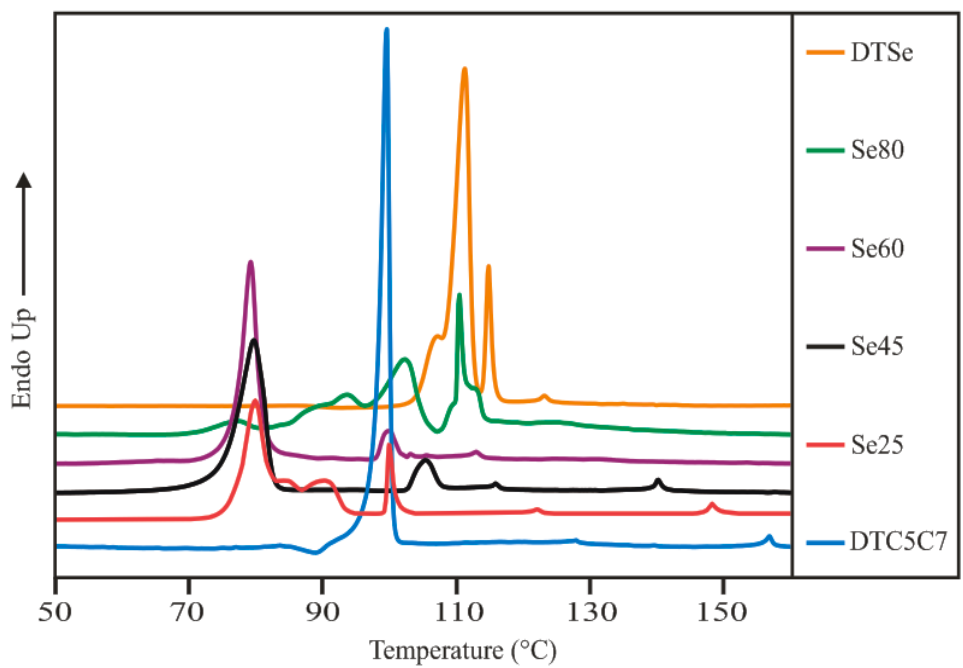

Figure S3 – First heating DSC curves of DTC5C7, DTSe and mixtures of the two compounds. The scanning rate was 10°C/min in all cases.

Table S1 – Position of temperature maxima

| Sample | C → SmC (°C) | C → $N_{tb}$ (°C) | SmC → $N_{tb}$ (°C) | SmC → N (°C) | $N_{tb}$ → N (°C) | N → Iso (°C) |
|---|---|---|---|---|---|---|
| DTC5C7 | --- | 99.5 | --- | --- | 127.8 | 156.8 |
| Se25 | 80 | --- | 99.8 | --- | 122.3 | 148.5 |
| Se45 | 79.5 | --- | 105.3 | --- | 115.8 | 140 |
| Se60 | 79 | --- | 99.8 | --- | 112.8 | 131 |
| Se80 | 102.1 | --- | --- | 110.3 | --- | 125.3 |
| DTSe | 111.2 | --- | --- | 114.8 | --- | 123.1 |

The temperatures corresponding to the DSC peak positions of DTC5C7, DTSe and mixtures of both compounds. C = Crystal, SmC = Smectic C, $N_{tb}$ = twist bend nematic, N = Nematic, Iso = Isotropic.

## 4.0 – X-ray Scattering of Se45

### 4.1 – X-ray Fluorescence of Se45

At photon energies close to the absorption (K) edge of the atom, electrons in the inner-most shells may be excited, which upon relaxation, may re-emit the absorbed photons with reduced energy. This is known as inelastic scattering or fluorescence. The fluorescence of Se45 was investigated on station I22 of the diamond light source in the energy range 12.608 – 12.708KeV, i.e. above and below the Se K-edge of 12.658KeV. Throughout the experiment the sample was held in a 1mm glass capillary. A 2D diffraction pattern was recorded in each 10eV increment of the energy range and converted to 1D by azimuthal integration. The change in background scattering, attributed to fluorescence, was determined from the baseline of the 1D diffraction patterns. These results are plotted in Figure S4 below for three different temperatures. At energies above and equal to the resonant energy, background scattering increases by a factor of ~2.

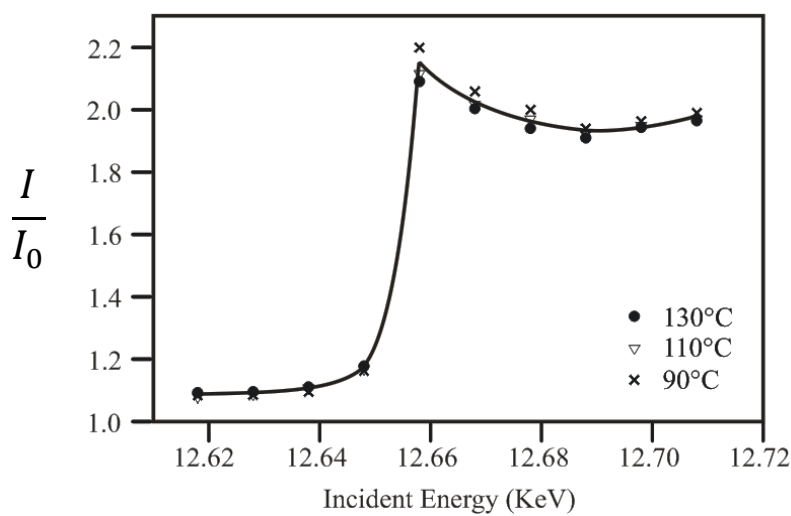

Figure S4 – The energy dependent background intensity created by fluorescent (inelastically scattered) photons in the range of $q$ = 0.3 – 3.7nm$^{-1}$. The quantity $I/I_0$ represents the scaling factor between the initial background intensity ($I_0$), measured at 12.608KeV, and the background intensity measured in at each 0.01KeV increment up to 12.708KeV.

### 4.2 – Resonant scattering features of Se45

The resonant X-ray scattering (RXS) investigation of Se45 was performed on beam-line I22 of the Diamond Light Source. The sample was held in a 1 mm glass capillary. The capillary was placed in a permanently magnetised cell such that the 1T magnetic field ran perpendicular to its length. The sample was heated and cooled using nitrogen gas (Cryojet HT by Oxford Instruments). Resonance of Se45 was firstly investigated at 105°C (N$_{tb}$ phase) in the energy range 12.608 – 12.708 keV, i.e. above and below the Se K-edge (12.658KeV). The sample to detector distance was 2.11m and the beam path was held under vacuum. Figure S5 shows 1D diffractograms recorded in each 10eV increment of the energy range. At the Se K-edge (12.658 keV) a resonant diffraction peak was observed at $q \sim$ 0.62 nm$^{-1}$, corresponding to a d-spacing ~10 nm.

The resonance effect was further investigated with temperature. The sample was heated and cooled using nitrogen gas (Cryojet HT by Oxford Instruments). A maximum temperature difference of 5°C was observed between nominal temperature and DSC peak position. As mentioned in the main text the resonance effect was observed only in the N$_{tb}$ phase and at the biphasic boundaries with the neighbouring N and SmC phases. Table S2 below summarises the key observations of temperature variation on the scattering maxima observed in Se45.

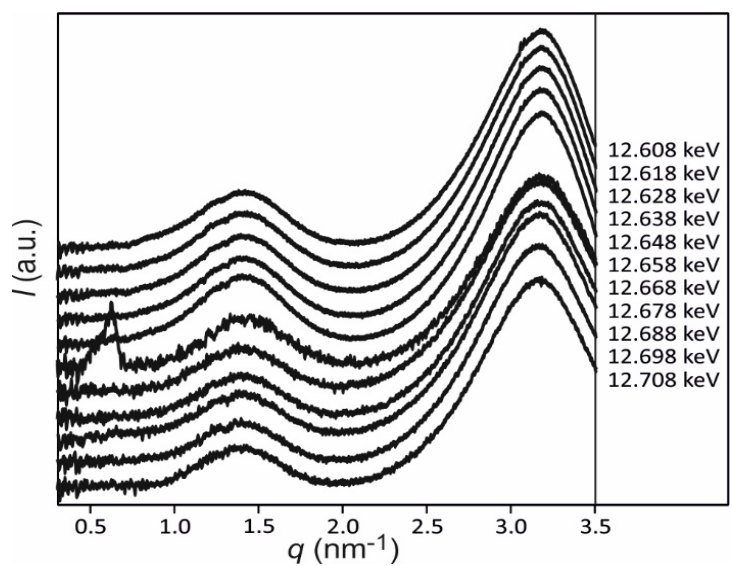

Figure S5 – The 1D diffractograms of Se45 recorded at 105°C, with incident beam energy changing from 12.608 – 12.708 keV. The resonant diffraction peak is observed at the Selenium absorption edge 12.658keV. The curves have been corrected for background scattering.

Table S2 – Summary of RXS data of Se45 during temperature scan.

| T (°C) | p (nm) | d1 (nm) | d2 (nm) | $\Delta q_{//p}$ (nm$^{-1}$) | $\Delta q_{//d2}$ (nm$^{-1}$) | d3 (nm) |
|---|---|---|---|---|---|---|
| 120 | - | 4.46 | 2.03 | - | 1.28 | 0.46 |
| 115 | 12.2 | 4.46 | 2.03 | 0.050 | 1.10 | 0.46 |
| 112 | 12.1 | 4.41 | 2.03 | 0.060 | 1.18 | 0.46 |
| 110 | 10.8 | 4.38 | 2.01 | 0.090 | 1.22 | 0.46 |
| 107 | 10.4 | 4.36 | 1.99 | 0.075 | 1.18 | 0.46 |
| 105 | 10.1 | 4.36 | 1.99 | 0.055 | 1.08 | 0.46 |
| 102 | 9.8 | 4.35 | 1.98/2.07* | 0.050 | 0.015 | 0.45 |
| 100 | 9.3 | 4.14 | 2.06 | 0.050 | 0.015 | 0.45 |
| 95 | - | 4.12 | 2.06 | - | 0.015 | 0.45 |
| 90 | - | 4.12 | 2.06 | - | 0.015 | 0.45 |

Development of SmC phase: rows at 102, 100, 95, 90 °C.

$p$ = helical pitch length, $d1$ and $d2$ respectively correspond to the average inter- and intra- molecular spacings determined from the non-resonant SAXS peaks. $d3$ is the d-spacing of the WAXS peak maximum, this is related to the lateral distance between molecules, which equals to 1.117 $d3$ (A. de Vries. Mol. Cryst. Liq. Cryst. **10**. 219 (1970).). The FWHM along the helical axis ($\Delta q_{//}$) of the peaks associated with $p$ and $d2$ are also provided. *Here the scattering maximum of the N$_{tb}$ phase, and the Bragg reflection of the SmC phase, are simultaneously observed. The larger of the two spacings corresponds to the SmC Bragg reflection.

## 4.3 – Non-resonant scattering features of Se45

The non-resonant small and wide angle scattering (SAXS and WAXS) features of Se45 were additionally investigated on station BM28 of the European Synchrotron Radiation Facility (ESRF). The sample was investigated in both transmission and grazing incidence setups. In the transmission setup the sample was sealed in a 1.0mm diameter glass capillary and placed in a custom built heating cell. The cell was in turn placed between the poles of a superconducting solenoid, which produced a magnetic field of 3T. The magnetically aligned WAXS and SAXS diffraction patterns were collected separately using a Mar165 CCD detector. The sample to detector distance was ~30cm for the WAXS patterns and ~1m for the SAXS patterns. In both instances a helium flushed flight tube was used to reduce air scattering. The WAXS patterns are shown in

Figure 3(e-g) of the main text, while the SAXS patterns are provided below in Figure S6. Similar to the evidence provided in the main text, alignment with the field worsens with temperature, until orientation is lost entirely in the SmC phase. The SAXS peaks can also be seen to shift away from the meridian (vertical axis in S6) in the $N_{tb}$ phase.

In the grazing incidence setup Se45 was melted onto a silicon substrate and cooled into the $N_{tb}$ phase (110°C). The resulting thin film was then sheared to produce homeotropic alignment of the helical axis. Homeotropic alignment persisted into SmC phase, with the layer normal perpendicular to the substrate surface. This can be seen in the grazing incidence x-ray diffraction (GIXRD) patterns of the $N_{tb}$ and SmC phases (S6). Homeotropic alignment is confirmed by the centring of the small angle peak(s) along the substrate surface normal (meridian).

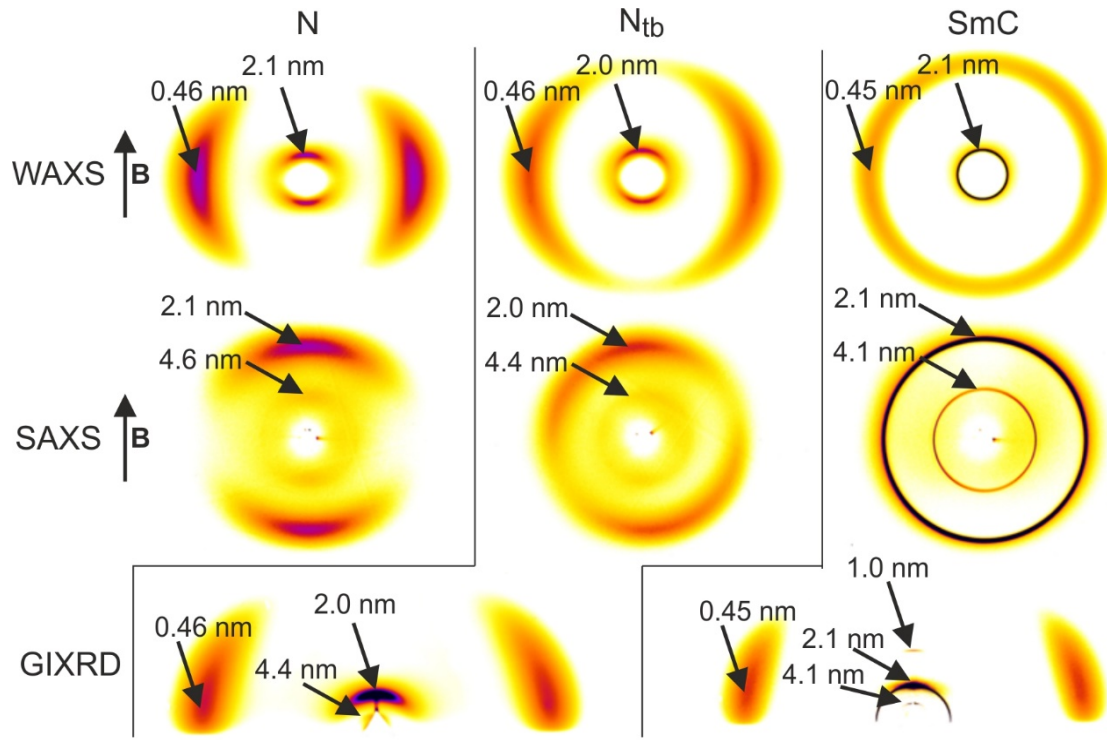

Figure S6 – Diffractograms of Se45, recorded far below the resonant energy. N: Nematic 140-116°C, $N_{tb}$: Twist Bend Nematic 115-106°C and SmC: Smectic-C 105-76°C. In the WAXS and SAXS patterns, the direction of the magnetic field is vertical. The GIXRD patterns show that better orientational order can be achieved in the $N_{tb}$ and SmC phases by shearing sample.

## 5.0 – Geometric calculations of the helical structure

Here we derive the geometric equations describing the behaviour of the molecules in the $N_{tb}$ phase. From these equations one may estimate the tilt angle of the mesogens and the bend angle of the dimers, using the experimentally determined values of the pitch length ($p$) and inter-mesogen height difference ($h$). As mentioned in the main text, $p$ is determined from the spacing of the resonant peak ($p$, Table S2), while $h$ is determined from the spacing of the outer-most non-resonant SAXS peak ($d2$, Table S2).

## 5.1 – Tilt angle of the mesogens $\theta$

Firstly the value of $h$ is linked to the helical contour length between the centres of two mesogens ($l$) by $h = l\cos(\theta)$. Here $\theta$ is the tilt angle between the mesogens and the helical axis $\mathbf{n_0}$. This relationship is shown diagrammatically in Figure S7 below. The average value of $l$ in Se45 can be determined from the molecular structures of DTC5C7 and DTSe. The spacer length, i.e. the distance between the last carbon on each of the two mesogens, is on average is 0.254nm x 4.5 = 1.14 nm. The length of the mesogen (three successive benzene rings) is ~2.4 nm. The average value of $l$ is therefore 1.14 + ( 2.4/2) = 2.34 nm. As the value of $l$ is assumed constant and the measured value of $h$ is almost invariant with temperature $\simeq$ 2nm, $\theta$ is calculated as 29±1° throughout the temperature range of the $N_{tb}$ phase.

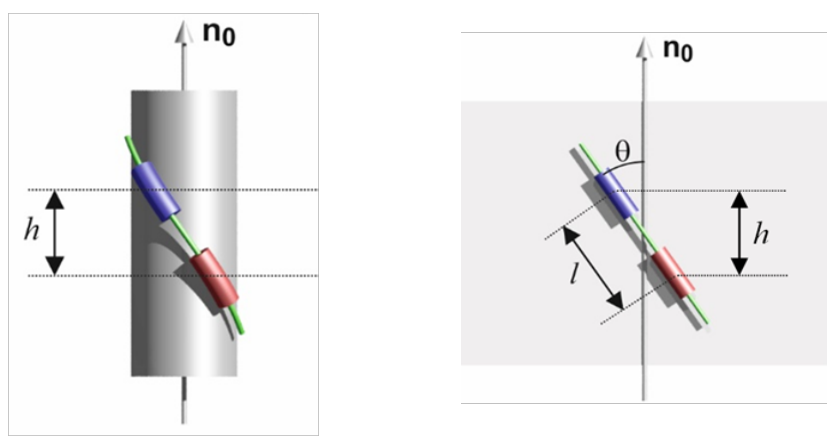

Figure S7 – Diagrammatic representation of the relationship between $l$, $h$, $\theta$ and the helical axis $\mathbf{n_0}$. On the left the molecule is wrapped around a cylinder and the distance between the two mesogens follows a helical contour. An unwrapped version of the cylinder is shown on the right. Here the molecule is straight (mesogens parallel).

## 5.2 - Bend angle of the dimers $\phi$

As mentioned in the main text, on progression along $\mathbf{n_0}$ each mesogen is rotated about the helical axis by an angle of $2\pi h/p$ with respect to the previous one. Here the rotation angle will be denoted by α. If each of the two rod-like mesogens of a dimer are respectively assigned unit vectors $\mathbf{n_1}$ and $\mathbf{n_2}$, the exterior angle between them ($\phi$) can be written:

$$\sin\frac{\phi}{2} = \sin\theta \sin\frac{\alpha}{2} \quad (Eqn. S1)$$

$\phi$ can be defined as the bend angle of the dimers. As $\sin(\theta)$ is approximately constant in the $N_{tb}$ phase, the bend angle is dependent only on the rotation angle α, which in turn is dependent on the pitch length $p$. As shown in Table S2, $p$ ranges between 9 – 12nm, indicating that the minimum and maximum values of the bend angle $\phi$ are 29 and 38° respectively.

## 5.3 – Tilt angle of the dimers β

An additional quantity, not discussed in the main text, is the effective tilt angle of the dimers with respect to the helical axis. Where the symbols have their previously defined meanings, the radius $r$ of the cylinder (hence the helix) can be defined from simple helical geometry as:

$$r = \frac{p \tan \theta}{2\pi} \qquad (Eqn.\,S2)$$

We may also define a chord length $w$, which represents the projected distance between two mesogens in the plane perpendicular to **n₀**.

$$w = 2r \sin\left(\frac{\alpha}{2}\right) = \frac{p \tan \theta}{\pi} \sin\left(\frac{\alpha}{2}\right) \qquad (Eqn.\,S3)$$

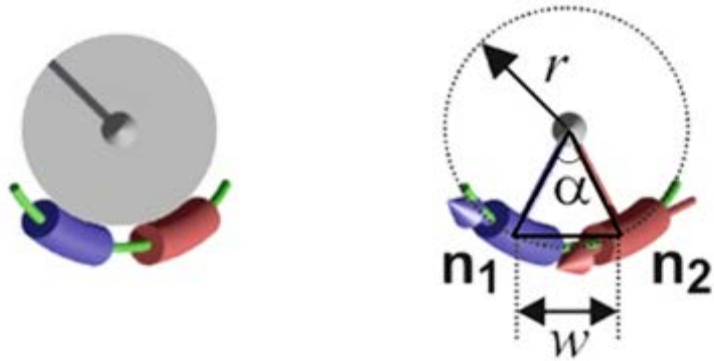

Figure S8 – Diagrammatic representation of the rotation angle between successive mesogens ($\alpha$), the unit vectors **n₁** and **n₂**, the radius of the helix ($r$) and the projected chord length $w$.

The tilt angle of the dimer $\beta$, which is the angle between the helical axis and the line connecting the centres of the two mesogens, can then be defined from simple geometry as:

$$\tan \beta = \frac{w}{h} = \tan \theta \, \frac{\sin(\frac{\alpha}{2})}{\alpha/2} \qquad (Eqn.\,S4)$$

The tilt angle $\beta$ as a function of the mesogen tilt angle $\theta$ and pitch is listed in Table S3.

Table S3 – The dimer tilt angle $\beta$ as a function of the mesogen tilt angle $\theta$ and helical pitch $p$.

| $\theta$\$p$ | 90nm | 100nm | 110nm | 120nm |
|---|---|---|---|---|
| 20° | 18.00° | 18.37° | 18.65° | 18.86° |
| 25° | 22.60° | 23.05° | 23.38° | 23.64° |
| 30° | 27.26° | 27.78° | 28.16° | 28.45° |
| 35° | 32.00° | 32.57° | 32.99° | 33.31° |
| 40° | 36.83° | 37.44° | 37.88° | 38.22° |
| 45° | 41.75° | 42.38° | 42.84° | 43.19° |

## 6.0 – Molecular modelling

We have carried out molecular modelling, in order to show how a smaller bend angle ($\phi$), between the two mesogenic arms of a dimer molecule, can be achieved through changing of the C-C torsion angles in the spacer. We have also estimated the energy cost associated with such changes of torsion angles in the spacer.

The modelling of the molecules is carried out using Materials Studio (BIOVIA). In the modelling, the conformations of the (-$C_7H_{14}$-) spacer are studied, starting from the all-trans conformation where all six torsion angles in the spacer are $\gamma = 180°$. Assuming that the spacer is always helical, the six torsion angles are kept to the same value $\gamma$, changing from 180° to 150° with 5° steps. The bend angle $\phi$ is measured for each torsion $\gamma$ (Figure S9) and listed in Table S4. The energies of the spacer in different conformations are calculated, using Universal Forcefield, and compared in Table S5. According to Table S5, the ideal torsion angle $\gamma$ should be around 160° as the experimentally measured bend angle $\phi$ from X-ray studies is around 29±1°. Even though the energy involved in such a conformational change from the all-trans state seems to be high, ~5.8 kcal/mol, it is an over-estimation as many other different conformations the spacer can adopt for the same bend angle are not explored, and such high energy conformational state can in fact be stabilized by the high entropy of the spacer, i.e. the various different conformations the spacer can take.

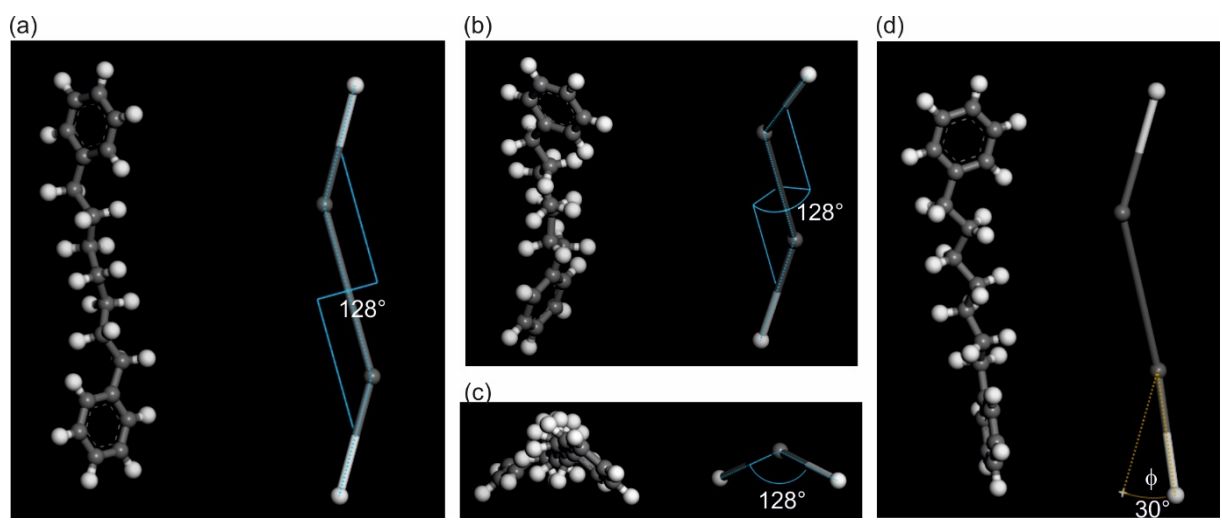

Figure S9. The molecular model of DTC7 with torsion angles in the spacer $\gamma = 160°$. For simplicity only the spacer and the two phenyl rings directly connected to the spacer are shown. To show the molecular geometry more clearly a rod model of the molecule is shown on the right hand side, where the C-C bond represents the spacer, and C-H bonds the mesogens. (a-c) The dihedral angle of the two mesogens around the spacer is measured to be 128°: (a) view when the spacer is parallel to the paper; (b) tilted view, (c) view along the spacer. (d) The bend angle $\phi$ of the dimer, i.e. the angle between the two mesogens, is measured to be 30°.

Table S4. The bend angle $\phi$ of a dimer as a function of torsion angles $\gamma$ adopted by the spacer.

| $\gamma$ (°) | $\phi$ (°) |
|---|---|
| 180 | 56.1 |
| 175 | 54.0 |
| 170 | 48.4 |
| 165 | 39.3 |
| 160 | 29.6 |
| 155 | 14.9 |
| 150 | 1.4 |

Table S5. Energy estimation of spacer (-$C_7H_{14}$-) with different torsion angles $\gamma$.

| $\gamma$ (°) | Total (kcal/mol) | Valence (kcal/mol) | Non-Bond (kcal/mol) |
|---|---|---|---|
| 180 | 3.997 | 1.248 | 2.750 |
| 175 | 4.359 | 1.463 | 2.896 |
| 170 | 5.444 | 2.096 | 3.348 |
| 165 | 7.250 | 3.106 | 4.144 |
| 160 | 9.755 | 4.423 | 5.332 |
| 155 | 12.941 | 5.952 | 6.989 |
| 150 | 16.880 | 7.604 | 9.276 |